\documentclass[pra,twocolumn,amsmath,amssymb,floatfix,reprint,footinbib,superscriptaddress,longbibliography,showkeys]{revtex4-2}
\usepackage{picinpar,graphicx}
\usepackage{braket}
\usepackage{amsmath}
\usepackage{graphicx}
\usepackage{epstopdf}
\usepackage{dcolumn}
\usepackage{graphicx}
\usepackage{mathrsfs}
\usepackage{mdwlist}
\usepackage{subfigure}
\usepackage{booktabs}
\usepackage{amsmath}
\usepackage{dsfont}
\usepackage{amstext}
\usepackage{amssymb}
\usepackage{amsbsy}
\usepackage{bbm}
\usepackage{amsthm}
\usepackage{graphicx}
\usepackage{textcomp}
\usepackage{multirow}
\usepackage{color}
\usepackage[colorlinks,citecolor=blue]{hyperref}
\definecolor{Dgreen}{RGB}{0, 100, 0}
\usepackage{url}
\usepackage[colorlinks]{hyperref}
\hypersetup{%
	plainpages=true,
	breaklinks=true,  
	hypertexnames=false,  
	pageanchor=true,
	colorlinks=true,
	linkcolor={blue},
	citecolor={red},
	urlcolor={blue},
	anchorcolor={black}
}

\usepackage{xcolor}
\usepackage[normalem]{ulem} 
\begin{document}

	\title{Neural network approach to mitigating intra-gate crosstalk in superconducting CZ gates}
	\author{Yiming Yu}
    \affiliation{Fujian Key Laboratory of Quantum Information and Quantum Optics, Fuzhou University, Fuzhou 350116, China}
	\affiliation{Department of Physics, Fuzhou University, Fuzhou, 350116, China}

	\author{Yexiong Zeng}
	\affiliation{Quantum Information Physics Theory Research Team, Center for Quantum Computing, RIKEN, Wako-shi, Saitama 351-0198, Japan}
	
	
	
	\author{Ye-Hong Chen}\thanks{yehong.chen@fzu.edu.cn}
    \affiliation{Fujian Key Laboratory of Quantum Information and Quantum Optics, Fuzhou University, Fuzhou 350116, China}%
	\affiliation{Department of Physics, Fuzhou University, Fuzhou, 350116, China}
	\affiliation{Quantum Information Physics Theory Research Team, Center for Quantum Computing, RIKEN, Wako-shi, Saitama 351-0198, Japan}
	
	\author{Franco Nori}
	\affiliation{Quantum Information Physics Theory Research Team, Center for Quantum Computing, RIKEN, Wako-shi, Saitama 351-0198, Japan}%
	\affiliation{Department of Physics, University of Michigan, Ann Arbor, Michigan 48109-1040, USA}
	
	\author{Yan Xia}\thanks{xia-208@163.com}
    \affiliation{Fujian Key Laboratory of Quantum Information and Quantum Optics, Fuzhou University, Fuzhou 350116, China}
	\affiliation{Department of Physics, Fuzhou University, Fuzhou, 350116, China}

	\date{\today}
	
	\begin{abstract}
		 The potential of quantum computing is fundamentally constrained by the inherent susceptibility of qubits to noise and crosstalk, particularly during multi-qubit gate operations. 
		 Existing strategies, such as hardware isolation and dynamical decoupling, face limitations in scalability, experimental feasibility, and robustness against complex noise sources. 
		 In this manuscript, we propose a physics-guided neural control (PGNC) framework to generate robust control pulses for superconducting transmon qubit systems, specifically targeting crosstalk mitigation. 
		 By combining a hardware aware parameterization with a Hamiltonian-informed objective that accounts for condition-dependent crosstalk distortions, PGNC steers the search toward smooth and physically realizable pulses while efficiently exploring high dimensional control landscapes.
		 Numerical simulations for the CZ gate demonstrate superior fidelity and pulse smoothness compared to a Krotov baseline under matched constraints. 
		 Taken together, the results show consistent and practically meaningful improvements in both nominal and perturbed conditions, with pronounced gains in worst-case fidelity, supporting PGNC as a viable route to robust control on near-term transmon devices.
	\end{abstract}
	
	\maketitle
	
\section{INTRODUCTION}
Quantum computing holds immense promise for solving classically intractable problems \cite{Cheng2023,Huang2020,wendin2017quantum,annurev:/content/journals/10.1146/annurev-conmatphys-031119-050605}, but its practical realization is fundamentally limited by qubit errors and especially by crosstalk in two-qubit gate operations \cite{PhysRevApplied.12.054023}.
Therefore, mitigating such crosstalk has become a pressing priority for building reliable and scalable quantum computers.
As demonstrated by Sarovar et al. \cite{Sarovar2020}, robust detection and characterization of crosstalk errors are crucial for benchmarking and improving quantum processors.
A key challenge is to implement high-fidelity single-qubit gates without inducing unwanted interactions in neighboring qubits \cite{Tanttu2024}. 
Multi-qubit gates typically require qubits to be spatially or spectrally proximate, which can result in control pulses unintentionally affecting nearby qubits. 
For example, a resonant pulse intended for one qubit may inadvertently drive transitions in a neighboring qubit \cite{pioro2008electrically}. 
Such crosstalk can cause unwanted entanglement or state mixing, severely degrading gate fidelity \cite{malekakhlagh2020first,rol2019fast,carleo2019machine}.

Crosstalk generally appears in two forms: quantum and classical \cite{Sarovar2020,zhao2022quantum,PhysRevApplied.12.064022}. 
Quantum crosstalk originates from intrinsic couplings between qubits, such as always-on interactions or deliberately engineered coupling elements \cite{PhysRevB.77.014510,deGroot2012}. 
These couplings—often dominated by ZZ interactions—shift the qubit energy levels and can cause unintended entanglement and state mixing, thereby corrupting the intended quantum state \cite{winick2021simulating,PhysRevLett.125.200504,PhysRevB.74.184504,PhysRevB.76.132513,PhysRevB.74.172505,PhysRevLett.96.067003}. 
Classical crosstalk, in contrast, arises from unintended electromagnetic coupling. Control signals intended for one qubit can leak into adjacent control lines, causing spurious operations on neighboring qubits.
This form of crosstalk is particularly challenging to mitigate due to its dependence on circuit design and operating frequencies \cite{Wang2022,Yang2024,Yang20241,PhysRevA.93.042307}. 
Regardless of form, crosstalk remains a major obstacle to achieving fault-tolerant quantum computing.
Therefore, mitigating crosstalk is crucial for improving the reliability and scalability of quantum computing.

Existing crosstalk-suppression techniques have several limitations. 
Hardware-based approaches, such as spacing out qubits or adding shielding, increase fabrication complexity and cost \cite{PhysRevApplied.12.054023,PhysRevX.11.021058,PhysRevApplied.22.044072,PhysRevResearch.6.013142}. 
However, such measures can also degrade the performance of individual qubits \cite{barends2014superconducting}. 
Although dynamical decoupling can counter certain noise sources \cite{bylander2011noise,8759071,tripathi2022suppression,PhysRevLett.133.033603,zhou2023quantum,PhysRevA.99.042327}, it is sensitive to the noise spectrum and requires carefully designed pulse sequences \cite{Li2022,hocker2014characterization}. 
Furthermore, iterative microwave-crosstalk calibration can substantially suppress residual leakage in small- to medium-scale systems \cite{yan2023calibration}.
However, the required calibration workload grows rapidly with system size and is sensitive to hardware drift and model mismatch, creating challenges for large-scale deployment.

Precisely manipulating quantum systems with carefully designed external fields is a core approach to realizing advanced quantum technologies \cite{brif2010control, rabitz2000whither, PhysRevApplied.18.064059,refId0,PhysRevX.3.041013}.
A key aspect of such control is pulse parameterization, which converts a continuous-time control problem into a tractable optimization problem.
This is achieved by discretizing control fields into a finite-dimensional parameter space \cite{chen2019fast, werschnik2007quantum, jirari2006quantum}. 
Optimization algorithms then search this space for pulse shapes that realize the desired operations with high fidelity \cite{Dong2010, shi2021two, kang2018pulse, chen2021shortcuts}. 
Oftentimes, pulse optimization must pursue multiple goals beyond just high fidelity. 
For example, pulses may be designed to counteract dissipation \cite{PhysRevLett.111.030405}, eliminating leakage to unwanted states \cite{PhysRevLett.106.190501,PhysRevA.84.022326}, or achieving robust control in specific quantum hardware like superconducting qubits \cite{PhysRevA.83.012308, PhysRevLett.102.090401}. 
Recently, learning-based control methods (e.g., neural network approaches) have emerged as powerful tools for discovering complex and robust control strategies \cite{norambuena2024physics, PhysRevLett.130.043604, PhysRevResearch.7.L012049}.
Related neural-network-based pulse synthesis for entangling gates under explicit crosstalk models has also been explored in other platforms, e.g., silicon spin qubits~\cite{Kanaar2024}.
Traditional optimal control methods such as GRAPE \cite{khaneja2005optimal}, DRAG \cite{motzoi2009simple,chen2016measuring} and Krotov \cite{PhysRevA.68.062308,10.21468/SciPostPhys.7.6.080,10.1063/1.3691827} can yield high-fidelity pulse solutions in theory. 
However, in practice they suffer from poor scaling with parameter dimensionality, susceptibility to local minima, and difficulty in simultaneously enforcing hardware constraints such as bandwidth and amplitude limits.
Furthermore, these methods often struggle to find robust solutions under complex crosstalk conditions \cite{soare2014experimental,rudinger2021experimental}. This challenge has prompted dedicated research into robust control strategies and learning-based pulse design methods \cite{Dong20150, Dong20160}.

To overcome these challenges, this work proposes a novel physics-guided neural control (PGNC) framework. 
This framework is a neural-network-based open-loop pulse generator that maps time $t$ and a crosstalk-condition vector $c$ to hardware-constrained waveforms, trained with a physics-based objective.
It enables efficient exploration of a high-dimensional control space by incorporating condition vector that encode different crosstalk (concurrent-drive) operating conditions.
By training on varying crosstalk conditions, a single network learns to generate optimal pulses that adapt to different crosstalk or detuning levels, rather than being fixed to one operating point \cite{norambuena2024physics,gungordu2022robust}.
By jointly optimizing multiple objectives (fidelity, leakage suppression, and pulse smoothness), the proposed framework more effectively reduces control errors in practice.
Unlike conventional gradient-based schemes, PGNC embeds the system model directly into the training process. 
This strategy steers the search toward smooth, experimentally realizable pulses with greater robustness against moderate crosstalk.
This study focuses on two-qubit gate operations, where the interplay between control fields and crosstalk is especially pronounced \cite{PhysRevB.74.184504}. These operations serve as an ideal testbed for evaluating the robustness and scalability of the proposed approach.

This paper is organized as follows. Sec.~\ref{s2} introduces the theoretical framework, including an open-system two-transmon model with condition-augmented Hamiltonians that capture both coherent (ZZ) and classical (cross-drive) crosstalk, as well as the hardware-aware pulse parameterization and training objectives. Sec.~\ref{s3} describes the numerical simulation setup and reports the results: the environment and parameters, PGNC training dynamics and learned waveforms, and a head-to-head comparison with a Krotov baseline under matched conditions (including fidelity distributions, robustness sweeps over crosstalk parameters, and resource usage analysis). Sec.~\ref{s4} discusses the mechanisms underlying the results, practical implications and limitations of the approach, and concludes with avenues for future work.


\section{Theoretical Framework} \label{s2}
This section outlines the theoretical framework for the study by specifying the quantum system, the noise model, the crosstalk interactions, and the resulting Hamiltonian. A system of $N$ superconducting transmon qubits is considered, a prevalent platform for quantum computing. Each qubit is characterized by its anharmonicity and control parameters that enable the manipulation of its quantum states. It is acknowledged that these qubits are susceptible to both quantum and classical noise, as well as various forms of crosstalk.

\subsection{Open-system two-qubit control model}\ \label{s2.1}

The device is modeled as a Markovian open quantum system evolving under the Lindblad master equation (natural units $\hbar=1$):
\begin{equation}
\begin{aligned}
\dot{\rho}(t)
&= -\,i\,[\,H(t),\,\rho(t)\,]
\\
&\quad + \sum_{k}\Big(L_k\,\rho(t)\,L_k^\dagger - \tfrac{1}{2}\{\,L_k^\dagger L_k,\,\rho(t)\,\}\Big),
\end{aligned}
\label{eq:lindblad}
\end{equation}
where $H(t)$ is the total Hamiltonian and $\{L_k\}$ the jump operators for
$T_1/T_2$-type channels; next, we use $T_{1,q}$ and $T_{2,q}$ to represent the energy relaxation time and coherence time of qubit $q$, respectively.

The Hamiltonian is decomposed into static and control contributions:
\begin{equation}
H(t)=H_{\mathrm{0}}+H_{\mathrm{ctrl}}(t).
\label{eq:Hsplit}
\end{equation}

We consider a tunable coupling architecture where the interaction is dynamically activated.
We work in the rotating frame of the qubit drive frequencies (and apply the rotating wave approximation), so the harmonic terms $\sum_q \omega_q n_q$ are removed.
Thus, the static Hamiltonian $H_0$ describes two independent anharmonic oscillators:
\begin{equation}
\begin{aligned}
H_{\mathrm{0}}
&= \frac{\alpha_1}{2}\,n_1\,(n_1-\mathbb{I})
+ \frac{\alpha_2}{2}\,n_2\,(n_2-\mathbb{I}).
\end{aligned}
\label{eq:Hdrift}
\end{equation}
Here, $\alpha_q (<0)$ represents the anharmonicity of qubit $q\in\{1,2\}$. The number operators $n_q = b_q^\dagger b_q$ describe the qubit excitations, where $b_q$ and $b_q^\dagger$ are the single-qubit ladder operators. We lift these operators to two-qubit space, defining $b_1 = b \otimes \mathbb{I}_{n_L}$ and $b_2 = \mathbb{I}_{n_L} \otimes b$ for qubits 1 and 2, respectively, where $n_L$ is the single-transmon truncation dimension, and $\mathbb{I}_{n_L}$ denotes the $n_L\times n_L$ identity.

The control Hamiltonian employs in-phase/quadrature drives and instantaneous detuning, and the time-dependent coupling modulated by the tunable coupler:
\begin{equation}
	\begin{aligned}
		H_{\mathrm{ctrl}}(t) =& \sum_{q=1}^{2} \left[ \Omega_{xq}(t) H_{qI} + \Omega_{yq}(t) H_{qQ} + \delta_q(t)\,n_q \right]\\ &+ J_{zz}(t) n_{1} n_{2},
	\end{aligned}
	\label{eq:Hctrl}
\end{equation}
with control generators $H_{qI}=\tfrac{1}{2}(b_q+b_q^\dagger)$ and $H_{qQ}=\tfrac{1}{2i}(b_q^\dagger-b_q)$. Here $\Omega_{xq}(t)$ and $\Omega_{yq}(t)$ are the $I/Q$ drive envelopes for qubit $q$, $\delta_q(t)$ is the instantaneous detuning in the rotating frame, and $J_{zz}(t)$ is the dynamically tunable effective $ZZ$ coupling strength.

Dissipation uses amplitude damping and pure dephasing channels with
$L_{1}^{(T_1)}=\sqrt{1/T_{1,1}}\,b_1$, 
$L_{2}^{(T_1)}=\sqrt{1/T_{1,2}}\,b_2$,
$L_{1}^{(\phi)}=\sqrt{\gamma_{\phi,1}}\,n_1$,
$L_{2}^{(\phi)}=\sqrt{\gamma_{\phi,2}}\,n_2$,
where the pure-dephasing rates satisfy
\begin{equation}
\gamma_{\phi,q}=\max\!\Big(0,\ \tfrac{1}{T_{2,q}}-\tfrac{1}{2T_{1,q}}\Big).
\label{eq:dephrate}
\end{equation}
Here the truncation $\gamma_{\phi,q}\ge 0$ enforces complete positivity.

\subsection{Crosstalk and condition-augmented model}\label{s2.2}

This section extends the Sec.~\ref{s2.1} model with a crosstalk-condition augmented description of coherent and classical crosstalk under concurrent operations, covering crosstalk-condition dependent $ZZ$ shifts, cross-drive mixing, detuning biases, and narrowband bleed-through \cite{PhysRevApplied.12.054023,zhao2022quantum}.

We introduce a dimensionless crosstalk-condition vector $c\in[-1,1]^3$ to parameterize concurrent-drive operating conditions,
\begin{equation}
c=\big[c_I,\ c_Q,\ c_f\big]^{\top},
\end{equation}
where $c_I$ and $c_Q$ are the normalized neighbor-tone amplitudes on the $I/Q$ channels and $c_f$ is a normalized carrier-offset tag.
The components of $c$ indicate concurrent-drive perturbations (e.g., increased cross-drive leakage and larger carrier-frequency offsets), and enter the condition-augmented crosstalk model by modulating the effective control parameters in the Hamiltonian (such as unintended drive amplitudes and the effective $ZZ$ shift).

\begin{figure*}[htbp] 
    \centering
    \includegraphics[scale=0.52]{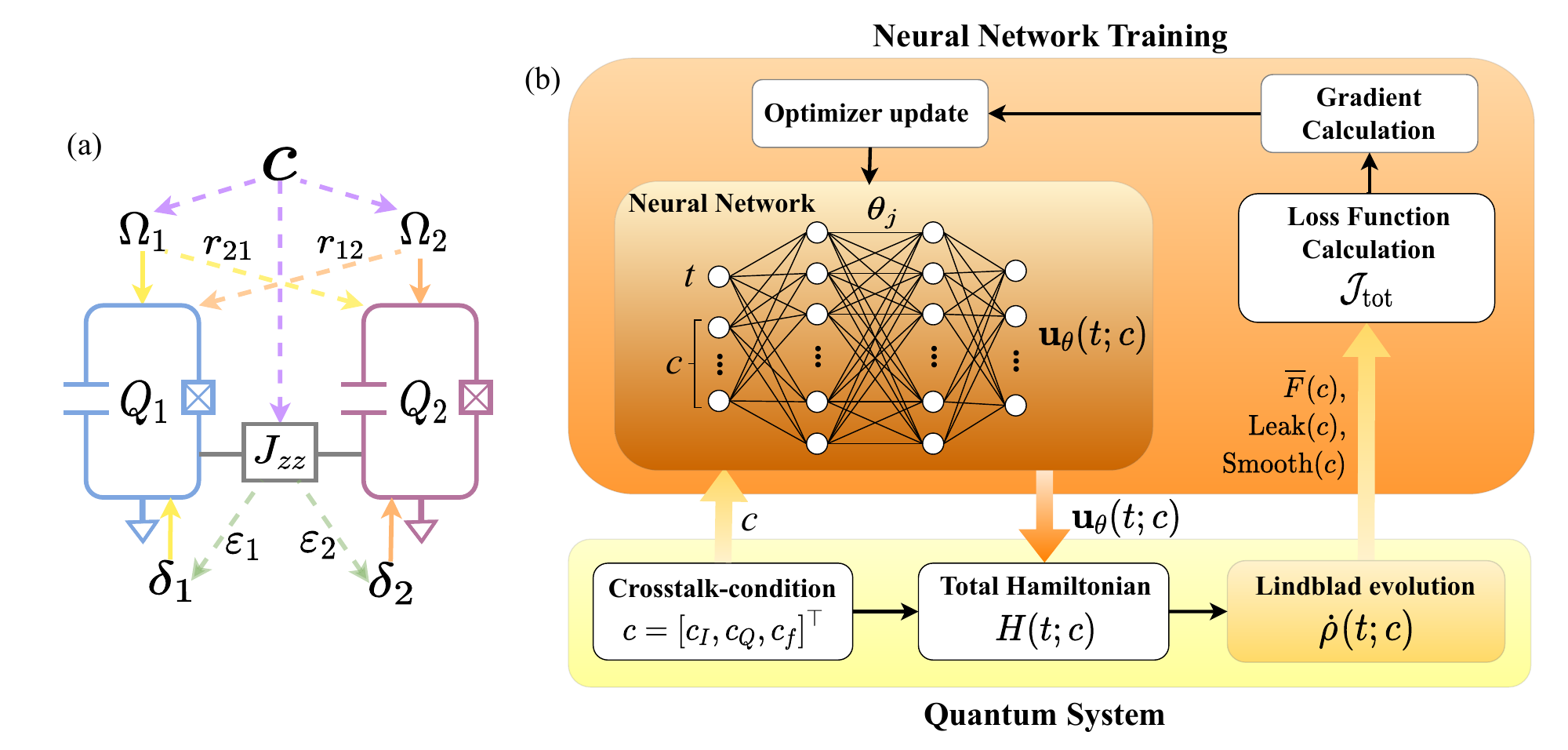} 
    \caption{
		(a) Conceptual illustration of coherent and crosstalk in a two-transmon system.
		Each transmon qubit $Q_q$ is controlled by local microwave I/Q envelopes $\Omega_q$ together with an instantaneous detuning term $\delta_q$, while coupled through an effective interaction $J_{zz}$, $q\in\{1,2\}$.
		Solid arrows denote the intended drives applied directly to the qubits, whereas dashed arrows denote corsstalk-induced perturbations to the effective control parameters.
		Classical cross-drive leakage is captured by the mixing coefficients $r_{12}$ and $r_{21}$.
		A normalized crosstalk-condition vector $c$ summarizes concurrent-drive settings and induces condition-dependent biases in the effective parameters, e.g., $J_{zz}^{\mathrm{eff}}(t;c)$ and $r_{\mathrm{eff}}(c)$ (Eqs.~\ref{eq:b_def} and \ref{eq:G_def}).
		The coefficients $\varepsilon_{1,2}$ denote parasitic $Z$-shifts correlated with activating $J_{zz}(t)$ (Eq.~\ref{eq:coupler_zshift}). 
		(b) Conditioned neural-network training loop for quantum control.
		Given $(t,c)$, the neural network outputs hardware-constrained waveforms $u_{\theta}(t;c)$, which define the condition-augmented Hamiltonian $H(t;c)$.
		The open-system dynamics are simulated via Lindblad evolution, from which we compute $\overline{F}(c)$, $\mathrm{Leak}(c)$, and $\mathrm{Smooth}(c)$ and form the aggregated training objective $\mathcal{J}_{\mathrm{tot}}$.
		Gradients are obtained by automatic differentiation and used by an optimizer to update $\theta$ until convergence; after training, fidelities are reported on a held-out evaluation set.}
\label{flow}
\end{figure*}

These effects are modeled to first order in $c$, which is appropriate for moderate concurrent-drive conditions (e.g., $|c_i|\lesssim 0.25$) where condition-induced AC-Stark shifts and drive-mediated couplings vary smoothly with the operating point; under stronger drives, higher-order corrections or a re-identification of sensitivity coefficients may be required.
The condition dimensionality can be expanded to include additional nuisance parameters when needed, but we adopt the above three-component form throughout this work. 
Importantly, the first-order dependence is applied to the condition-to-parameter map (effective $ZZ$ shift, drive-mixing coefficient, and detuning biases), i.e., a locally identifiable Jacobian around the nominal operating point, rather than assuming the underlying crosstalk mechanisms are globally linear.
The total Hamiltonian with crosstalk-condition reads \(H(t;c)=H_{\mathrm{0}}+H_{\mathrm{ctrl}}(t;c)\). Figure~\ref{flow}(a) summarizes the condition-augmented mechanisms used to construct $H(t;c)$.

To capture slow, condition-dependent distortions induced by concurrent activity, we adopt a first-order affine model
\begin{equation}
b(c)=b_0+\mathbf{G}\,c,
\label{eq:b_affine}
\end{equation}
The parameter vector $b(c)\in\mathbb{R}^4$ collects the effective quantities entering our simulator:
\begin{equation}
b(c)=
\begin{bmatrix}
J_{zz}^{\mathrm{eff}}(t;c)\\
r_{\mathrm{eff}}(c)\\
\Delta\delta_1(c)\\
\Delta\delta_2(c)
\end{bmatrix},
\qquad
b_0=
\begin{bmatrix}
J_{zz}(t)\\
r\\
0\\
0
\end{bmatrix}.
\label{eq:b_def}
\end{equation}
Here $J_{zz}^{\mathrm{eff}}(t;c)$ denotes the condition-biased effective $ZZ$ coupling, and $r_{\mathrm{eff}}(c)$ is the condition-biased cross-drive leakage coefficient.
The quantities $\Delta\delta_q(c)$ ($q\in\{1,2\}$) represent condition-induced quasi-static detuning biases.
The zero entries in $b_0$ state that, in the absence of concurrent activity ($c=\mathbf{0}$), the condition-induced detuning offsets vanish,
i.e., $\Delta\delta_q(\mathbf{0})=0$.
The specific physical roles and operating mechanisms of these parameters are discussed in detail later in this subsection.

The sensitivity matrix $\mathbf{G}\in\mathbb{R}^{4\times 3}$ collects absolute first-order sensitivities,
\begin{equation}
\mathbf{G}=
\begin{bmatrix}
g_J^\top\\
g_r^\top\\
g_{\Delta,1}^\top\\
g_{\Delta,2}^\top
\end{bmatrix},
\qquad
g_J,g_r,g_{\Delta,1},g_{\Delta,2}\in\mathbb{R}^{3},
\label{eq:G_def}
\end{equation}
so that each row induces the corresponding shift via $g_\bullet^\top c$.
In each case, $g_\bullet$ quantifies the first-order sensitivity of the corresponding effective parameter to the condition vector $c$.
Here $g_J$ has the same units as $J_{zz}$, $g_r$ is dimensionless (same as $r$), and $g_{\Delta,q}$ has the same units as $\delta_q$.

This first-order model is adopted as a simple and transparent parameterization to present the condition dependence in a compact form and to keep the benchmarking attribution-clean.
In practice, the PGNC pipeline does not rely on linearity: the condition-to-parameter map can be re-identified from calibration data and replaced by higher-order expansions,
nonlinear regressors, or piecewise/local models whenever the device exhibits stronger nonlinearity or operates outside the moderate range considered here.

Classical crosstalk is first modeled as linear mixing of the physical in-phase/quadrature envelopes. Collect the drives as $\boldsymbol{\Omega}(t)=[\,\Omega_{x1}(t),\ \Omega_{y1}(t),\ \Omega_{x2}(t),\ \Omega_{y2}(t)\,]^{\top}$ and let
\begin{equation}
\begin{aligned}
	\boldsymbol{\Omega}^{\mathrm{lin}}(t)=\mathbf{M}(r)\,\boldsymbol{\Omega}(t),
	\qquad
	\mathbf{M}(r)=\begin{bmatrix}
	1&0&r_{12}&0\\
	0&1&0&r_{12}\\
	r_{21}&0&1&0\\
	0&r_{21}&0&1
	\end{bmatrix},
\end{aligned}
\label{eq:mixing}
\end{equation}
where $r_{12}$ and $r_{21}$ quantify cross-drive leakage. For simplicity, we assume symmetric cross-drive leakage, i.e., $r_{12}=r_{21}=r$. The framework readily accommodates asymmetric crosstalk by replacing $\mathbf{M}(r)$ with distinct $r_{12}$ and $r_{21}$ entries without altering the rest of the formulation. 
We further assume that linear cross-coupling predominantly occurs between like quadratures ($I$-to-$I$, $Q$-to-$Q$), and neglect direct $I$-to-$Q$ mixing in $\mathbf{M}(r)$. Such cross-quadrature effects, if present, can be incorporated by adding the corresponding off-diagonal blocks.

Parasitic coherent bleed-through from concurrent tones is modeled as a condition-modulated narrowband injection, windowed by a bounded flattop envelope $\mathrm{env}(t)$ satisfying $\mathrm{env}(0)=\mathrm{env}(T)=0$:
\begin{equation}
\begin{aligned}
\Delta\boldsymbol{\Omega}(t;c)
&= \mathbf{D}\,c\;\mathrm{env}(t)\,\sin\!\big(\Delta f(c)\,t\big),\\
\Delta f(c)
&= \kappa^{\!\top} c ,
\end{aligned}
\label{eq:level2}
\end{equation}
where $\Delta\boldsymbol{\Omega}(t;c)=[\Delta\Omega_{x1},\Delta\Omega_{y1},\Delta\Omega_{x2},\Delta\Omega_{y2}]^{\!\top}$ stacks the injected $I/Q$ components on qubits 1 and 2.
This term is added to the nominal drive envelopes, i.e.,
$\Omega_{xq}^{\mathrm{eff}}(t;c)=\Omega_{xq}(t)+\Delta\Omega_{xq}(t;c)$ and
$\Omega_{yq}^{\mathrm{eff}}(t;c)=\Omega_{yq}(t)+\Delta\Omega_{yq}(t;c)$.
The matrix $\mathbf{D}\in\mathbb{R}^{4\times 3}$ maps the condition $c$ to the per-channel injection amplitudes, so that each channel receives a sinusoidal bleed-through whose strength is linear in $c$.
The offset frequency is likewise taken to be condition dependent, $\Delta f(c)=\kappa^\top c$ with $\kappa\in\mathbb{R}^3$, capturing a first-order shift of the dominant interference tone under concurrent activity.

In practice, $\mathbf{D}$ and $\kappa$ are calibration-facing nuisance parameters that summarize the dominant coherent spur under concurrent tones. They can be identified from calibration data by measuring the induced I/Q leakage and fitting it to Eq.~\ref{eq:level2}. In this work we treat them as fixed coefficients shared across training and evaluation, but they could also be learned or periodically re-identified to track drift.

In tunable-coupler operation, modulating the effective interaction $J_{zz}(t)$ is typically accompanied by parasitic single-qubit frequency shifts on the computational qubits~\cite{PhysRevApplied.12.054023,PhysRevApplied.12.064022,PhysRevX.11.021058}.
To capture this hardware-mapped nonideality, we model an additional detuning component proportional to the activated coupling waveform:
\begin{equation}
	\Delta\delta^{(Z)}_q(t)=\varepsilon_q\,J_{zz}(t),
\qquad q\in\{1,2\},
\label{eq:coupler_zshift}
\end{equation}
where $\varepsilon_q$ are dimensionless sensitivity coefficients that quantify the strength of the coupling-to-$Z$ leakage on each qubit.

Combining linear mixing, condition-induced injection and dynamically coupling-induced detuning gives the effective drives and the condition-augmented control Hamiltonian 
\begin{equation}
\begin{aligned}
	\boldsymbol{\Omega}^{\mathrm{eff}}(t;c)
	= \mathbf{M}[r_{\mathrm{eff}}(c)]\,\boldsymbol{\Omega}(t)
	+ \Delta\boldsymbol{\Omega}(t;c).
\end{aligned}
\label{eq:Omega-eff}
\end{equation}
\begin{equation}
\begin{aligned}
&H_{\mathrm{ctrl}}(t;c)
= \sum_{q=1}^2\Big[
\Omega^{\mathrm{eff}}_{xq}(t;c)\,H_{qI}
+ \Omega^{\mathrm{eff}}_{yq}(t;c)\,H_{qQ}\\
&\qquad\qquad
+ \delta_q(t;c)\,n_q
\Big]+ J_{zz}^{\text{eff}}(t;c) n_{1} n_{2},
\end{aligned}
\label{eq:Hctrl-ctx}
\end{equation}
with condition-biased detuning $\delta_q(t;c) = \delta_q(t) + \Delta\delta_q(c) +
\Delta\delta^{(Z)}_q(t), q\in\{1,2\}.$

\subsection{Control parameterization}\label{s2.3}

This subsection specifies the mapping from a neural generator to hardware-realizable waveforms that directly produce $\Omega_{xq}(t;c)$, $\Omega_{yq}(t;c)$, $\delta_q(t;c)$, and $J_{zz}(t;c)$ for $q\in\{1,2\}$. Here $t\in[0,T]$ is time within a gate of total duration $T>0$. The learned policy produces hardware-feasible waveforms; endpoint and bandwidth feasibility are enforced structurally by the envelope and saturation maps below. For clarity, Fig.~\ref{flow}(b) summarizes the end-to-end pipeline, and we next detail the control parameterization.

Time is normalized as $\tau=t/T\in[0,1]$. To represent smooth, band-limited temporal structure while allowing condition dependence, we embed $(t,c)$ through a $K$-order Fourier feature map:
\begin{equation}
\begin{aligned}
\phi(t;c)
&=
\Big[
\,\tau,\;
\{\sin(2\pi k \tau),\ \cos(2\pi k \tau)\}_{k=1}^{K},\;
c
\Big]^{\!\top},
\end{aligned}
\label{eq:phi}
\end{equation}
with $K\ge 1$ (default $K=4$). The low-order harmonics capture the dominant in-gate
temporal modes, $\tau$ conveys slow trends and edge behavior, and $c$ injects the
scenario dependence at first use.

The feature vector is passed through a two-layer multilayer perceptron (MLP), i.e., a fully connected feedforward neural network, with parameters $\theta$ to produce a length-$7$ parameter vector.
\begin{equation}
\begin{aligned}
p(t;c)
&=
\mathrm{MLP}_{\theta}\!\big(\phi(t;c)\big)
\in\mathbb{R}^{7},
\\
p(t;c)
&=
\big[
p_{x1}(t;c),\,
p_{y1}(t;c),\,
p_{\delta 1}(t;c),\,
p_{x2}(t;c),\,\\
&~~~~~p_{y2}(t;c),\,
p_{\delta 2}(t;c),\,
p_{J}(t;c)
\big]^{\!\top}.
\end{aligned}
\label{eq:p_latent7}
\end{equation}
Each component $p_{\bullet}(t;c)$ is a channel-wise pre-saturation parameter.

Endpoint conditions and hardware feasibility are enforced via a common analytic flattop envelope,
\begin{equation}
\begin{aligned}
\mathrm{env}(t)
&=\frac{1}{\tanh\!\big(\tfrac{s}{4}\big)}
\Big[
\tanh\!\Big(\frac{s\,t}{4T}\Big)
-
\tanh\!\Big(\frac{s\,(t-T)}{4T}\Big)
\Big]\\
&\quad - 1.
\end{aligned}
\label{eq:env}
\end{equation}
The analytic flattop envelope is smooth on $[0,T]$, satisfies $\mathrm{env}(0)=\mathrm{env}(T)=0$, and remains approximately unity in the interior; the prefactor $1/\tanh(s/4)$ normalizes the plateau, while the terminal “$-1$” enforces zero endpoints. The shape parameter $s>0$ controls the rise/fall steepness (larger $s$ narrows the transition region but increases out-of-band spectral content). Consequently, the envelope enforces endpoint and amplitude constraints without post-processing and remains compatible with gradient-based training.

The latent-to-physical mapping is applied channel-wise for each qubit $q\in\{1,2\}$:
\begin{equation}
\begin{aligned}
\Omega_{xq}(t;c)
&=
\Omega_{\max}\,
\mathrm{env}(t)\,
\tanh\!\big(p_{xq}(t;c)\big),
\\[3pt]
\Omega_{yq}(t;c)
&=
\Omega_{\max}\,
\mathrm{env}(t)\,
\tanh\!\big(p_{yq}(t;c)\big),
\\[3pt]
\delta_{q}(t;c)
&=
\Delta_{\max}\,
\mathrm{env}(t)\,
\tanh\!\big(p_{\delta q}(t;c)\big),
\\[3pt]
J_{zz}(t;c)
&=
J_{\max}\,
\mathrm{env}(t)\,
\tanh\!\big(p_{J}(t;c)\big).
\end{aligned}
\label{eq:phys-map}
\end{equation}
Here $\tanh(\cdot)$ smoothly bounds each channel to $(-1,1)$ before scaling by $\Omega_{\max}$, $\Delta_{\max}$, or $J_{\max}$.
The shared envelope $\mathrm{env}(t)$ enforces zero endpoints and controlled edges for all channels.

For later use, we collect these channels into the control waveform vector
\begin{equation}
	\begin{aligned}
	\mathbf{u}(t;c)=
	\big[&
	\Omega_{x1}(t;c),\ \Omega_{y1}(t;c),\ \Omega_{x2}(t;c),\ \Omega_{y2}(t;c),\\
	&\delta_{1}(t;c),\ \delta_{2}(t;c),\ J_{zz}(t;c)
	\big]^{\!\top}.
	\end{aligned}
	\label{eq:u_def}
\end{equation}

\subsection{Optimization objective and evaluation protocol}\label{s2.4}
The evolution follows Eq.~\ref{eq:lindblad} with the condition-augmented Hamiltonian
$H(t;c)$.
The target two-qubit gate $U_{\mathrm{tar}}\in\mathrm{U}(4)$ is embedded into the full Hilbert space via $\mathcal{E}(\cdot)$, acting as identity on all leakage levels. 
Here $\mathcal{E}(\cdot)$ denotes the canonical embedding from the computational two-qubit space into the truncated multi-level Hilbert space.

Consider a set of pure two-qubit inputs $\{\ket{\psi_m}\}_{m=1}^{M}$ composed of the $16$ computational/$\pm$ product states together with Haar-random draws on the two-qubit space. Each state is embedded as $\ket{\Psi_m}=\mathcal{E}(\ket{\psi_m})$ and initialized as $\rho_m(0)=\ket{\Psi_m}\!\bra{\Psi_m}$. For a given condition $c$, time evolution produces $\rho_m(T;c)$. The corresponding target state is
\begin{equation}
\ket{\Psi_m^{\mathrm{tar}}}
=
\mathcal{E}\!\big(U_{\mathrm{tar}}\ket{\psi_m}\big).
\label{eq:target-embed}
\end{equation}
The sample fidelity is
\begin{equation}
\begin{aligned}
F_m(c)
&=
\bra{\Psi_m^{\mathrm{tar}}}\,\rho_m(T;c)\,\ket{\Psi_m^{\mathrm{tar}}}
\\
&=
\mathrm{Tr}\!\Big[\,
\rho_m(T;c)\;
\ket{\Psi_m^{\mathrm{tar}}}\!\bra{\Psi_m^{\mathrm{tar}}}
\Big],
\end{aligned}
\label{eq:fid-sample}
\end{equation}
and the average $\overline{F}(c)=\tfrac{1}{M}\sum_{m=1}^{M}F_m(c)$ serves as the core objective surrogate, i.e., a Monte Carlo estimate of average gate fidelity under open-system evolution, without reconstructing a process matrix. This Monte Carlo estimate of the average gate fidelity avoids explicit process reconstruction and is sufficiently accurate for comparing controllers under open-system evolution at practical computational cost.

Leakage out of the computational subspace is penalized via
\begin{equation}
\mathrm{Leak}(c)
=
\frac{1}{M}\sum_{m=1}^{M}
\Big(
1-\mathrm{Tr}\!\big[P_{\mathrm{comp}}\,\rho_m(T;c)\big]
\Big),
\label{eq:leak}
\end{equation}
where $P_{\mathrm{comp}}$ projects onto $\{|00\rangle,|01\rangle,|10\rangle,|11\rangle\}$
(for a ladder truncation, $P_{\mathrm{comp}}=\mathbb{I}_4\oplus \mathbf{0}$). Temporal roughness is controlled by an $H^1$-type smoothness penalty on the physical waveform vector
\begin{equation}
\begin{aligned}
\mathrm{Smooth}(c)
&=
\frac{1}{T}\int_{0}^{T}
\big\|
\partial_t\,\mathbf{u}(t;c)
\big\|_{2}^{2}\,dt.
\end{aligned}
\label{eq:smooth}
\end{equation}
Here, $\mathbf{u}(t;c)$ is defined in Eq.~\ref{eq:u_def}.
It is evaluated on an equally spaced grid using first-order differences. Endpoint and amplitude constraints are enforced structurally by the envelope and saturation maps in Eq.~\ref{eq:env}. The $\mathrm{Smooth}(c)$ is a proxy regularizer, not a hard bandwidth constraint: it discourages step-to-step jitter and excessive high-frequency content. If needed, hard bandwidth limits can be enforced via band-limited parameterizations (e.g., Fourier) or by low-pass filtering through a fixed/identified transfer function.

Let $\mathcal{C}=\{c^{(1)},\ldots,c^{(S)}\}$ denote the covered set of conditions to enumerate the concurrent/neighbor scenarios of interest, so that training and evaluation aggregate performance across all relevant operating conditions rather than a single nominal case. The per-condition objective is
\begin{equation}
\begin{aligned}
\mathcal{J}\!\big(c^{(s)}\big)
&= \big(1-\overline{F}(c^{(s)})\big)\\
&\quad +\, w_{\mathrm{leak}}\ \mathrm{Leak}(c^{(s)})\\
&\quad +\, w_{\mathrm{smooth}}\ \mathrm{Smooth}(c^{(s)}),
\end{aligned}
\label{eq:J-per-context}
\end{equation}
with positive weights $w_{\mathrm{leak}}$ and $w_{\mathrm{smooth}}$. 
In practice, $w_{\mathrm{smooth}}$ can be treated as a calibration knob: it is increased until the learned waveforms are compatible with the expected hardware bandwidth/transfer-function limitations, while preserving the best-achievable fidelity.

The weights are chosen to reflect practical priorities while keeping the penalty terms comparable in scale. Concretely, we set $w_{\mathrm{leak}}$ to suppress $\mathrm{Leak}(c)$ to the target level without noticeably degrading $\overline{F}(c)$, and set $w_{\mathrm{smooth}}$ to the smallest value that removes high-frequency jitter while preserving the best-achievable fidelity. The same weights are used for all simulations, and we verified that the main performance trends are insensitive to moderate variations of these weights.

The global objective aggregates across $\mathcal{C}$:
\begin{equation}
\mathcal{J}_{\mathrm{tot}}
=
\mathrm{Agg}\Big(\,\{\mathcal{J}(c^{(s)})\}_{s=1}^{S}\,\Big),
\label{eq:J-total}
\end{equation}
where $\mathrm{Agg}$ defaults to the arithmetic mean (balanced performance), and can be replaced by $\max$ or $\mathrm{CVaR}_\alpha$ to emphasize worst-case behavior. 
Here $\mathrm{CVaR}_\alpha$ denotes the conditional value-at-risk at level $\alpha\in(0,1]$, emphasizing the lower-tail (worst-case) conditions.

The evaluation protocol mirrors the objective: for each condition we compute $\{F_m(c)\}$ and summarize performance with standard distributional metrics and robustness sweeps over key physical parameters; full procedures and plots are detailed in Sec.~\ref{s3}.


\section{Numerical Simulations and Results}\label{s3}
\subsection{Environment and Parameter Settings}\label{s3.1}

We model two transmons truncated to three levels each, with transition frequencies $\omega_{1}/2\pi=4.380$\,GHz and $\omega_{2}/2\pi=4.614$\,GHz, anharmonicities $\alpha_{1}/2\pi=-240$\,MHz, $\alpha_{2}/2\pi=-243$\, and baseline crosstalk coefficient $r=0.05$ \cite{10.1063/1.5089550,annurev-conmatphys-031119-050605}. Decoherence follows a Lindblad master equation with $T_{1}=70~\mu$s and $T_{2}=80~\mu$s. Gate synthesis targets a CZ operation over $T=50$\,ns, discretized into $N=50$ uniform steps~\cite{rol2019fast,PhysRevX.11.021058}.

We encode concurrent-operation condition in a 3D vector $c=[c_I,c_Q,c_f]^\top\in[-1,1]^3$.
At level~1, we parameterize the condition dependence with the affine map in Eqs.~\ref{eq:b_affine}--\ref{eq:G_def}.
In our numerical setting we use $g_r=(0,0,0.2)$,
$g_J/2\pi=(0,0,2)\,\mathrm{MHz}$, and
$g_{\Delta,1}/2\pi=(3,0,-0.5)\,\mathrm{MHz}$,
$g_{\Delta,2}/2\pi=(0,3,-0.5)\,\mathrm{MHz}$.
This choice reflects a simplified separation where the dominant $r$ and $J_{zz}$ biases are tagged by the carrier-offset component $c_f$, while the quasi-static detuning offsets are taken to scale mainly with the concurrent-tone amplitudes $(c_I,c_Q)$.
Level~2 parasitic injections are modeled as a condition-tagged sinusoid with offset frequency $\Delta f(c)=\kappa^\top c$ (with a nominal scale of $5$~MHz), windowed by the flattop envelope in Eq.~\ref{eq:env}.

The PGNC controller is realized as a Fourier-featured MLP with $K=4$ harmonics and two hidden layers of 64 $\tanh$ units, producing 7 latents that are mapped to hardware-feasible waveforms via the shared envelope and a $\tanh$ bounding map. 
PGNC is trained with random condition augmentation in the three-dimensional condition space: we sample $c=[c_I,c_Q,c_f]^\top$ from the hyper-rectangle $c_I\in[0,0.25]$, $c_Q\in[0,0.25]$, and $c_f\in[-0.25,0]$. 
For each training run, we first draw a candidate pool of size $24$ uniformly from this region and then select $3$ conditions without replacement for training; the nominal condition $c_0=(0,0,0)$ is always included. 
Training runs for 400 epochs with Adam (initial step size $3\times 10^{-3}$), cosine decay, and global-norm clipping, minimizing a weighted sum of infidelity, leakage, and smoothness ($w_{\mathrm{leak}}=0.05$, $w_{\mathrm{smooth}}=0.01$). 
We report average fidelities using an independent set of $N_{\mathrm{Haar}}=512$ embedded Haar-random states.

As numerical optimal-control baselines, we use both Krotov and GRAPE for comparison.
In addition, we implemented a physically calibrated CZ baseline \cite{PhysRevA.96.022330} (Gaussian single-qubit rotations with virtual-$Z$ phase updates and an interaction segment generated by $J_{zz}(t)$) and evaluated it under the same condition-augmented open-system simulator. In this setting, it attains substantially lower fidelity ($\sim0.95$) than the optimal-control baselines($\sim0.999$), so we report it in text but omit it from the main overlay plots for readability.
Krotov starts from sin-modulated controls under the shared flattop envelope (rise time $10$ ns), updates only within $[0.1T,0.9T]$, and uses penalty factors $\lambda_{a}^{(1)}=\lambda_{a}^{(2)}=1$ and $\lambda_{\delta}^{(1)}=\lambda_{\delta}^{(2)}=2$; condition-induced biases and parasitic terms are treated as fixed (non-optimized) additions, and only the seven physical controls are updated.
GRAPE optimizes the same seven channels using a piecewise-linear waveform on the $N=50$ grid, mapped to hardware-feasible amplitudes via the shared envelope and a $\tanh$ saturator, and minimizes the same objective as PGNC using Adam.
Unless otherwise stated, all other settings (Hamiltonian model, state ensembles, and numerical tolerances) are identical to those used for PGNC.

All simulations use NumPy/JAX/Diffrax/Optax/Flax for differentiable control and
QuTiP/Krotov for propagation and optimal control; Diffrax’s adaptive Dopri5 solver with a PID controller and \texttt{RecursiveCheckpointAdjoint} (rel./abs.\ tol.\ $10^{-7}$) is used
throughout.

\begin{table}[t]
	\centering
	\caption{Key device and simulation parameters. Frequencies are quoted as angular rates; divisions by $2\pi$ are shown for readability.}
	\begin{tabular}{lcc}
	\hline
	Quantity & Value & Notes \\
	\hline
	$\omega_1/2\pi$ (GHz) & 4.380 & Transmon~1 transition \\
	$\omega_2/2\pi$ (GHz) & 4.614 & Transmon~2 transition \\
	$\alpha_1/2\pi$ (MHz) & $-240$ & Anharmonicity (T1-limited regime) \\
	$\alpha_2/2\pi$ (MHz) & $-243$ & Anharmonicity \\
	$r$ (—) & $0.05$ & Cross-drive ratio (nominal) \\
	$T_1, T_2$ ($\mu$s) & $70,\,80$ & Lindblad channels \\
	Gate time $T$ (ns) & 50 & with $N=50$ steps \\
	$n_L$ (levels) & 3 & Truncation per transmon \\
	\hline
	\end{tabular}
	\label{tab:params}
	\end{table}

\subsection{Training Dynamics and Learned Waveforms}\label{s3.2}

\begin{figure}[htbp] 
    \centering
    \includegraphics[scale = 0.50]{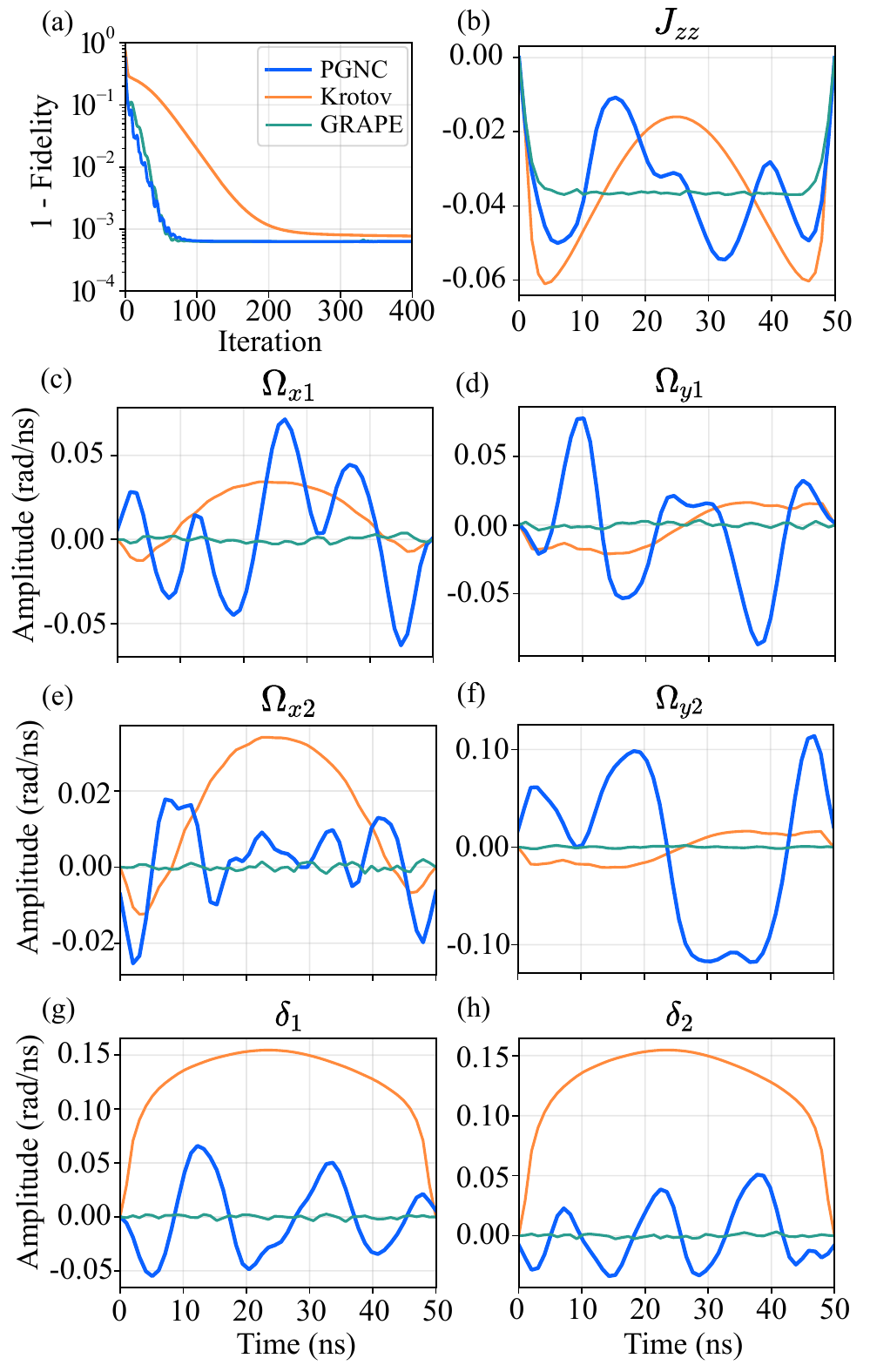} 
    \caption{Training dynamics and learned control waveforms for a two-qubit gate comparing PGNC, Krotov, and GRAPE algorithms. (a) Trace of the infidelity ($1 - \mathcal{F}$) versus training iteration on a logarithmic scale. (b-h) Learned hardware-feasible waveforms for the two qubits, plotted over the full gate window $T = 50$ ns: dynamic ZZ coupling ($J_{zz}$) (b), in-phase drives $\Omega_{x1}$ (c) and $\Omega_{x2}$ (e), quadrature drives $\Omega_{y1}$ (d) and $\Omega_{y2}$ (f), and detunings $\delta_{1}$ (g) and $\delta_{2}$ (h). Amplitudes and couplings are reported in rad/ns.}
    \label{F3}
\end{figure}

\begin{figure*}[t] 
    \centering
    \includegraphics[scale = 0.6]{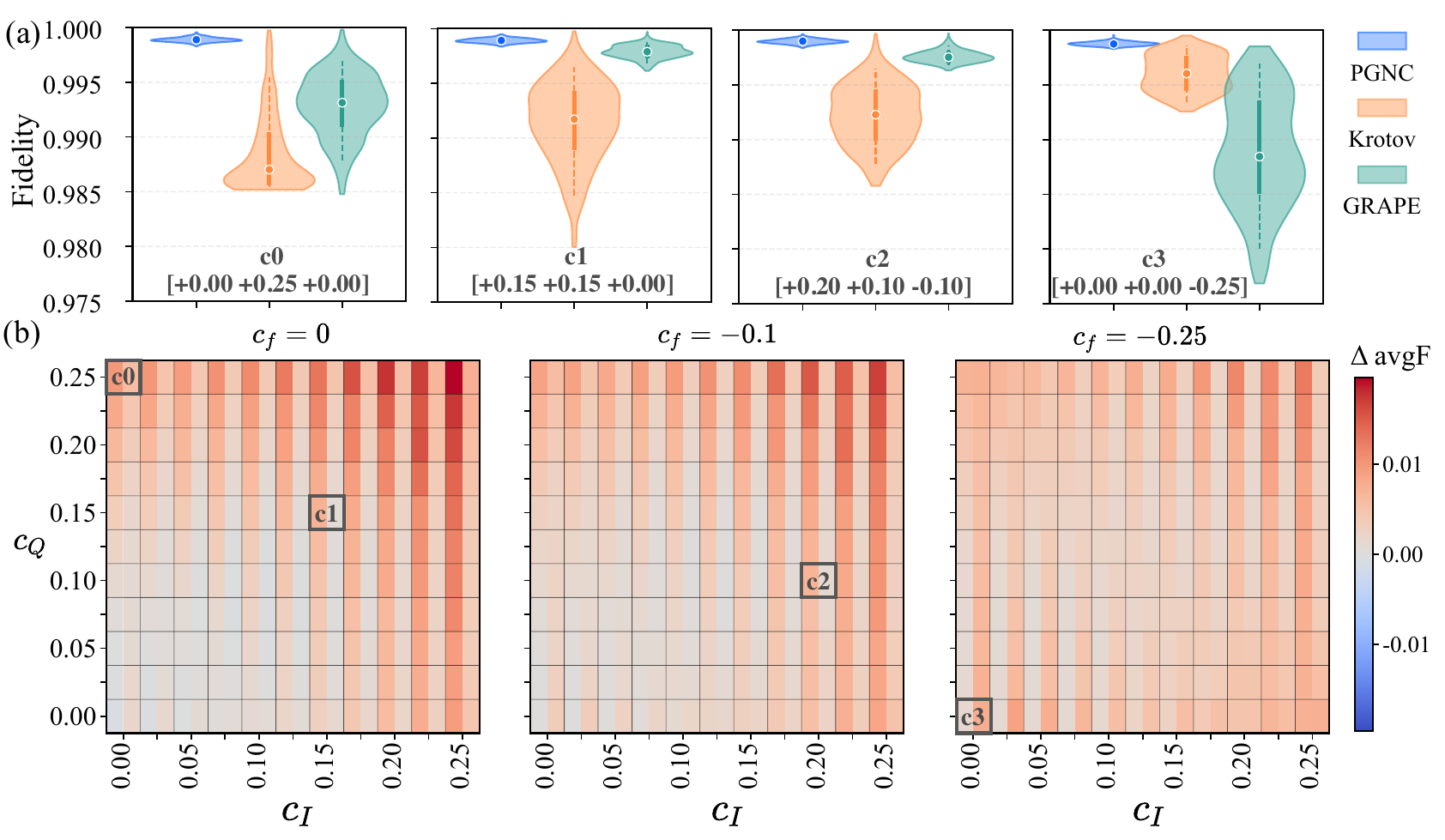} 
    \caption{Discrete-condition comparison and off-grid generalization scan.
			(a) Fidelity distributions (violin plots) for four representative conditions $c_0$--$c_3$ (annotated in each panel), evaluated on an ensemble of 128 Haar-random two-qubit input states, comparing PGNC, Krotov, and GRAPE.
			PGNC generates conditioned pulses by taking $c$ as an input at inference time, whereas Krotov and GRAPE are used as static baseline methods that output fixed waveforms under the chosen optimization setup and are then evaluated across different $c$.
			(b) Off-grid scan over $(c_I,c_Q)$ for three fixed carrier-offset tags $c_f\in\{0,-0.1,-0.25\}$.
			Each square panel contains two heatmaps: the left half shows the average-fidelity difference $\Delta\mathrm{avgF}=\mathrm{avgF}_{\mathrm{PGNC}}-\mathrm{avgF}_{\mathrm{Krotov}}$, and the right half shows $\Delta\mathrm{avgF}=\mathrm{avgF}_{\mathrm{PGNC}}-\mathrm{avgF}_{\mathrm{GRAPE}}$ (see color bar).
			Grey squares mark the locations of the four discrete conditions used in (a) on the corresponding scan planes.}
    \label{F4}
\end{figure*}

Figure~\ref{F3} summarizes the optimization dynamics and the learned control waveforms for the tunable-coupling CZ model introduced in Secs.~\ref{s2.1}--\ref{s2.2}, and compares PGNC against two standard optimal-control baselines (Krotov and GRAPE) under the same hardware constraints and time window.

Figure~\ref{F3}(a) shows the optimization trace in terms of the infidelity $1-F$, where $F$ is evaluated on the computational subspace using the same state ensemble as described in Sec.~\ref{s3.1}. Both PGNC and GRAPE rapidly reduce the infidelity and then plateau at the $10^{-3}$ level, whereas Krotov converges more slowly and reaches a slightly higher residual infidelity within the shown iteration budget. This behavior is consistent with the fact that PGNC and GRAPE directly optimize a parameterized waveform representation, while Krotov iteratively updates the continuous-time controls through its functional gradient with step-size regularization.

Figures~\ref{F3}(b--h) compare representative learned waveforms over the gate duration $T=50~\mathrm{ns}$. Panel (b) shows the tunable interaction $J_{zz}(t)$: Krotov produces a smooth, single-lobe waveform, while PGNC learns a more structured shape with alternating curvature within the allowed envelope, and GRAPE yields an almost time-independent profile within the same bounds. Panels (c--f) present the transverse drives $\Omega_{xq}(t)$ and $\Omega_{yq}(t)$. A salient feature of the PGNC solution is the stronger use of the $y$-quadrature on qubit 2 together with nontrivial modulations on the other quadratures, indicating a multi-channel cooperative strategy rather than relying on a single dominant knob. In contrast, the Krotov solution exhibits comparatively smooth, slowly varying quadrature components, and GRAPE keeps the transverse drives close to zero over most of the window. Finally, panels (g--h) show the instantaneous detunings $\delta_1(t)$ and $\delta_2(t)$: Krotov uses large positive detuning plateaus with a flattop-like profile, whereas PGNC and GRAPE operate with substantially smaller detuning excursions.

Regarding GRAPE, Figs.~\ref{F3}(b--h) indicate that the optimized solution is strongly biased toward an ``interaction-only'' realization of the CZ gate. The transverse drives remain close to zero for most of the gate window, and the entangling phase is generated predominantly by the coupling waveform $J_{zz}(t)$. While this is a valid theoretical mechanism in a tunable-coupler architecture, it acts effectively as a $J_{zz}$-dominant baseline that underutilizes the available microwave degrees of freedom.
In contrast, PGNC exhibits a fundamentally different multi-channel control strategy. Instead of relying on a single dominant knob or requiring large detuning plateaus like Krotov, PGNC actively exploits small but nonzero transverse components (e.g., a strongly modulated $\Omega_{y2}(t)$) alongside the tunable coupling. Taken together, Fig.~\ref{F3} demonstrates that under the same hardware amplitude constraints, PGNC avoids the local optima of purely $J_{zz}$-driven solutions and efficiently reaches a comparable or lower infidelity regime by distributing the modulation burden across all available control channels.

\subsection{Condition Generalization and Robustness}\label{s3.3}

We benchmark PGNC against two established optimal-control baselines, Krotov and GRAPE, under an identical open-system simulator, identical gate time $T$, identical control channels and bounds, and the same fidelity evaluation protocol described in Sec.~\ref{s3.1}. The key distinction is the role of the condition input. While the condition $c$ can be used to instantiate the Hamiltonian during evaluation for all methods, it is not a learnable conditioning variable for conventional static optimal control: a Krotov/GRAPE run outputs a single waveform, and providing $c$ does not induce a condition-dependent policy. In contrast, PGNC explicitly learns a conditional mapping $c \mapsto u_\theta(t;c)$ and can therefore adapt its waveform to different crosstalk conditions at inference time. This setting enables a direct comparison between a static optimal solution (Krotov/GRAPE) and a dynamic adaptive strategy (PGNC) in their response to perturbations.

Figure~\ref{F4} provides a unified view of how the three controllers behave across the condition space, combining discrete-condition distributions and an off-grid scan. 
Figure~\ref{F4}(a) compares fidelity distributions at four representative conditions $c0$--$c3$ [marked by the colored boxes in Fig.~\ref{F4}(b)]. 
In the nominal/near-nominal regimes ($c0$--$c2$), all methods achieve high fidelities, but PGNC consistently produces the tightest distributions, indicating fewer low-fidelity outliers under the same physical constraints. 
The contrast is most pronounced at $c_3$, where GRAPE exhibits a substantial degradation and a widened distribution, while PGNC remains near-unity and Krotov degrades only mildly. 
Beyond these four points, Fig.~\ref{F4}(b) scans $(c_I,c_Q)$ at three fixed values of $c_f$ and reports the average-fidelity gaps, where the left (right) halves of each panel show $\Delta \overline{F}=\overline{F}_{\mathrm{PGNC}}-\overline{F}_{\mathrm{Krotov}}$ ($\overline{F}_{\mathrm{PGNC}}-\overline{F}_{\mathrm{GRAPE}}$). 
Across the scanned region, the gaps are predominantly positive, and the baselines degrade more visibly as $c_f$ moves away from zero, consistent with a stronger impact of frequency-offset-related coherent terms in the Hamiltonian.

The off-grid maps show that the performance gaps are not uniform across the condition plane. The gaps do not increase monotonically with the distance from $c=\mathbf{0}$; instead, they vary across $(c_I,c_Q)$ and change noticeably as $c_f$ is varied. Accordingly, the three methods remain close near $c=\mathbf{0}$, while more off-nominal regions exhibit larger, $c_f$-dependent differences.
The representative condition in (a) are consistent with the surrounding neighborhoods in (b), indicating that the observed trends are not isolated to four selected points but persist over a continuous range of operating conditions.
These trends align with the intended operating mode of PGNC. During offline training, PGNC learns a conditional mapping from the condition vector $c$ to control waveforms, so when the crosstalk condition changes it can dynamically reshape $u_\theta(t;c)$---re-balancing the use of $J_{zz}(t)$, transverse quadratures, and detuning components---to counteract condition-dependent coherent errors without requiring a new optimization run for each $c$. 
Consequently, PGNC improves not only the average performance but also suppresses the lower tail of the fidelity distribution in off-nominal conditions, which is critical for reliable operation under fluctuating crosstalk conditions.

\begin{figure}[t] 
    \centering
    \includegraphics[scale = 0.6]{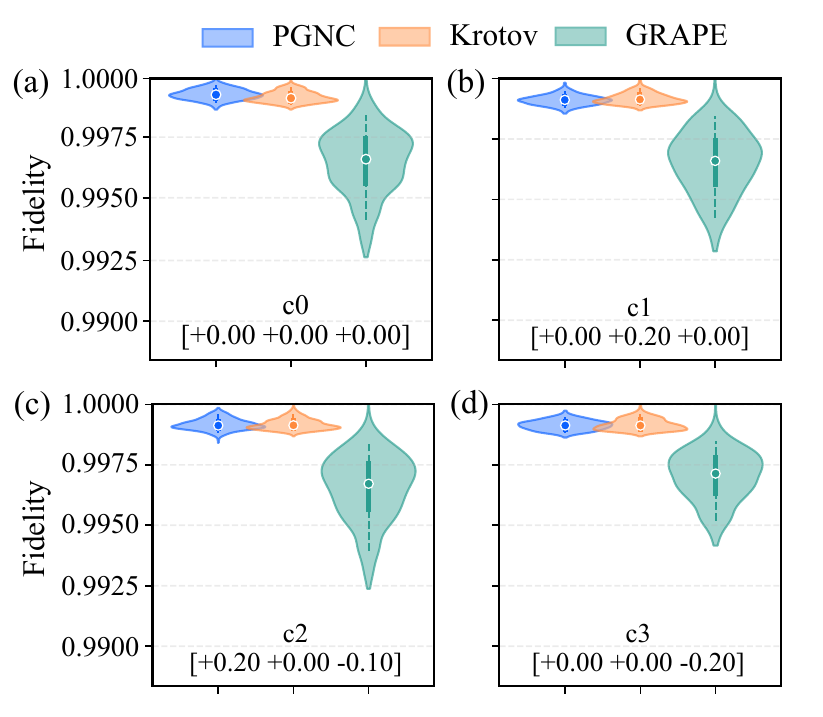} 
	\caption{Per-condition optimized CZ-gate fidelity distributions under representative crosstalk conditions.
    (a--d) Violin plots of the final gate fidelity obtained by independently optimizing PGNC, Krotov, and GRAPE for four fixed conditions $c0$--$c3$.
    All methods are evaluated under matched amplitude and smoothness constraints and the same optimization budget.
    The violin width indicates the distribution over repeated runs on an ensemble of 128 Haar-random two-qubit input states}
    \label{F5}
\end{figure}

Figure~\ref{F5} further benchmarks the three approaches under a per-condition protocol, where each method is allowed to specialize its optimization to a specific crosstalk condition. For each $c_k$, Krotov and GRAPE are re-optimized independently to produce condition-specific waveforms, whereas PGNC uses a single network trained offline and adapts only by conditioning on $c$ at inference time. Even under this more favorable setting for the baselines, PGNC remains competitive and consistently yields the tightest fidelity distributions across $c_0$--$c_3$, indicating fewer low-fidelity outliers. By contrast, GRAPE exhibits systematically broader spreads and a longer lower tail, suggesting stronger state-dependent residual coherent errors despite per-condition optimization; this is consistent with GRAPE favoring an interaction-dominant solution that can become less stable once the condition-dependent drive-mixing, injected components, and detuning-shift terms are non-negligible. Fig.~\ref{F5} shows that PGNC’s advantage is not merely averaging across conditions during training, but reflects a robust conditioned control strategy that preserves high-fidelity performance with reduced variance under realistic crosstalk nonidealities.

Overall, the combined evidence from Figs.~\ref{F4}--\ref{F5} supports the intended use-case of PGNC: a single offline-trained, conditioned controller that can generate high-fidelity pulses across a range of crosstalk conditions without requiring costly per-condition re-optimization.

\begin{figure}[t] 
    \centering
    \includegraphics[scale = 0.44]{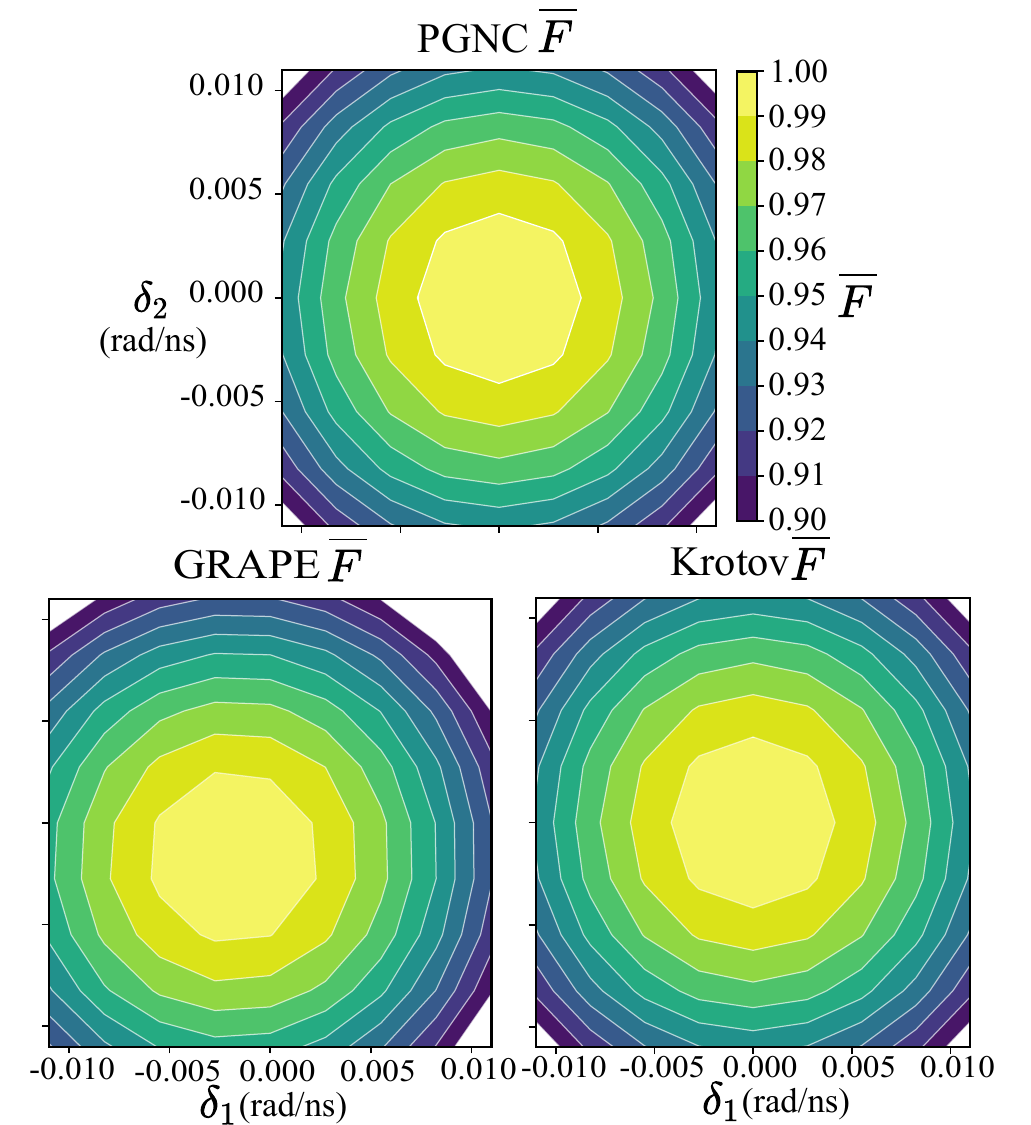} 
    \caption{
		Robustness to two-qubit detuning drifts.
		Contour maps of the average gate fidelity $\overline{F}$ as a function of static detuning offsets $(\delta_1,\delta_2)$ applied to the two qubits during the gate.
		The top panel shows PGNC, while the bottom panels show GRAPE and Krotov under the same physical model, constraints, and evaluation protocol.
		For each grid point $(\delta_1,\delta_2)$, we shift the detuning controls as
		$\delta_q(t)\mapsto \delta_q(t)+\delta_q$ ($q\in\{1,2\}$) and re-evaluate the gate fidelity on an ensemble of input states (the same evaluation set used throughout the paper), reporting the resulting average $\overline{F}$. The white central region indicates parameter pairs for which $\overline{F}>0.99$ (values above the color scale).
		}
    \label{F6}
\end{figure}

Figure~\ref{F6} evaluates robustness to quasi-static detuning offsets on the two qubits by scanning constant shifts $(\delta_1,\delta_2)$ applied throughout the gate at the nominal condition $c=\mathbf{0}$, i.e., all condition-dependent terms are fixed to their baseline values and only the detuning offsets are varied.
The contour maps report the resulting average gate fidelity $\overline{F}$ under identical constraints and the same evaluation set as in the main benchmarks.
All three methods exhibit a high-fidelity plateau around the nominal point, with the central white region indicating $\overline{F}>0.99$ (values above the plotted color scale).
Moving away from the origin, the fidelity gradually decreases and the level sets become anisotropic, indicating that the sensitivity depends on the direction in the $(\delta_1,\delta_2)$ plane rather than only on $|\delta_1|$ and $|\delta_2|$.
This detuning-sweep therefore provides a complementary robustness view to the discrete-condition tests, visualizing how each controller’s performance degrades under two-qubit frequency miscalibration.

\section{Discussion and Conclusion}\label{s4}

In this work, we proposed and validated a PGNC framework for mitigating intra-gate crosstalk in superconducting two-qubit gates. Numerical simulations show that, compared to conventional optimal-control baselines optimized at a single nominal condition, PGNC generates control pulses with higher fidelity and improved robustness across a range of crosstalk conditions, with pronounced gains in worst-case performance.

A central observation is that PGNC remains competitive not only under perturbed conditions but also at the nominal operating point $c_0$. This advantage can be attributed to a combination of optimization and representation effects. In particular, multi-condition training acts as an implicit regularizer: by optimizing one controller over diverse conditions, the learned solution tends to favor flatter regions of the control landscape, which are less sensitive to small perturbations while maintaining high fidelity.

Moreover, the PGNC parameterization (Fourier features and an MLP) biases the search toward smooth, hardware-feasible waveforms. As reflected by the tighter fidelity distributions in Fig.~\ref{F5}, smoother controls reduce unnecessary high-frequency content and help suppress leakage outside the computational subspace. We also observe nontrivial cross-channel coordination in the learned solutions (e.g., between $\Omega_{y2}$ and $\delta_1$), consistent with an AC-Stark-assisted compensation mechanism that achieves the target evolution through cooperative modulation rather than relying on a single dominant knob.

Importantly, the robustness improvement is not achieved by consuming additional control resources. Instead, PGNC preserves larger amplitude headroom in the available channels, which provides an effective buffer against parameter drifts and reduces the likelihood of abrupt performance collapse away from the nominal setting.

Beyond the intra-gate setting studied here, the proposed framework can be extended to inter-gate interference by augmenting the condition vector $c$ to encode the parallel-gate configuration around the target CZ. 
Concretely, for each simultaneously executed neighboring operation within a fixed local neighborhood, $c$ can include:
(i) timing descriptors (e.g., relative start time $\Delta t^{\mathrm{start}}$, overlap duration, or start/end offsets),
(ii) gate-/channel descriptors (e.g., whether the neighbor is a CZ/iSWAP/single-qubit drive and which control lines are active),
and (iii) low-dimensional pulse descriptors for the neighbor (e.g., a few calibrated template parameters such as amplitude/phase/detuning scalars, compact basis coefficients such as spline/Fourier coefficients, or summary features such as pulse energy and dominant spectral moments).
Training can then sample $c$ from a distribution of parallel-gate schedules (drawn from a device scheduler or synthetic workloads) and optimize the same conditioned mapping $c\mapsto \mathbf{u}(t;c)$. 
At inference time, the controller generates the target-gate pulse conditioned on the current parallel configuration, enabling adaptation to inter-gate interference without rerunning an optimizer for every configuration. 
Therefore, the same conditioned-control framework provides a direct path to addressing inter-gate crosstalk and can be further generalized to more complex concurrent-operation settings by enriching the condition descriptors.

In this work, our numerical demonstrations are restricted to a two-qubit effective model with an $n_L$-level truncation, and extending the simulator to large systems would indeed be challenging.
Note that PGNC does not circumvent the fundamental complexity of large-scale quantum simulation. 
Its practical value is to amortize calibration cost: after offline training on an effective model, generating pulses for new conditions reduces to a single forward pass conditioned on $c$, rather than per-condition re-optimization or extensive re-calibration. 
For larger devices, we do not require a full-chip Hamiltonian; instead, training can be based on an effective local gate-neighborhood model, since the dominant coherent interference is typically governed by nearby couplers/drives and a bounded set of concurrent operations. 
Under this locality assumption, the conditions can encode neighboring activity via low-dimensional, physically motivated descriptors (e.g., timing offsets and a few calibrated pulse parameters), keeping the learning problem tractable.

Despite the demonstrated potential of PGNC, several limitations remain. First, the present study is restricted to a two-qubit model, and scaling to larger devices will require locality-based modeling, reduced-order simulators, or hybrid hardware-in-the-loop updates. Second, as with all model-based approaches, performance depends on the fidelity of the assumed Hamiltonian and crosstalk model; integrating PGNC with online calibration and drift tracking is therefore an important direction. Finally, the same framework could be extended to other objectives beyond fidelity (e.g., noise-specific robustness or multi-objective formulations) for broader applicability.

Our approach is conceptually related to prior neural-network-based pulse synthesis that incorporates crosstalk and noise directly in the objective, such as the work of Kanaar on silicon spin qubits under charge noise and drive-induced crosstalk~\cite{Kanaar2024}. In contrast to optimizing a fixed pulse set for a given device instance, PGNC is a conditioned pulse generator that takes the condition vector $c$ as an explicit input and outputs condition-dependent waveforms $u(t;c)$ at inference time. Moreover, we introduce a calibration-facing, low-dimensional parameterization of concurrent-drive effects \{e.g., $b(c)=b_0+\mathbf{G}c$ and $\mathbf{M}[r_{\mathrm{eff}}(c)]$\} that supports attribution-clean benchmarking across conditions and a clear separation of static versus adaptive baselines under matched constraints.

Although some dressing effects (e.g., AC Stark shifts) are generally nonlinear in drive amplitude, using a first-order condition model is not contradictory: within a restricted operating window, the net parameter variation under moderate concurrent drives can be captured by a local linearization in $c$. If stronger nonlinearity becomes relevant, the same framework can incorporate higher-order terms or a low-dimensional nonlinear regressor for $\mathbf{b}(c)$ without changing the PGNC controller.

Overall, PGNC provides a practical route to crosstalk-resilient CZ-gate synthesis by amortizing optimization across conditions with a single conditioned controller, while remaining compatible with calibration-facing model updates when higher-order effects become relevant.

\section*{Acknowledgments}
 We would like to thank Yanming Che for helpful suggestions on the paper. Y.-H.C. was supported by the National Natural Science Foundation of China under Grant No. 12304390 and 12574386, the Fujian 100 Talents Program, and the Fujian Minjiang Scholar Program. Y.X. was supported by the National Natural Science Foundation of China under Grant No. 62471143, the Key Program of National Natural Science Foundation of Fujian Province under Grant No. 2024J02008, and the project from Fuzhou University under Grant No. JG2020001-2. F.N. is supported in part by: 
the Japan Science and Technology Agency (JST) [via the CREST Quantum Frontiers program Grant No. JPMJCR24I2, the Quantum Leap Flagship Program (Q-LEAP), and the Moonshot R\&D Grant Number JPMJMS2061].

\section*{Author contributions}
Yiming Yu conceived the study. Yiming Yu and Ye-Hong Chen developed the theoretical framework and designed the simulations. Yiming Yu implemented the code and performed the numerical experiments. Yiming Yu, Yexiong Zeng analyzed the results and wrote the manuscript. All authors discussed the results and commented on the manuscript.

\section*{Data availability}
The data that support the findings of this study are available from the corresponding author upon reasonable request.

\section*{Code availability}
Custom code used in this study is available from the corresponding author upon reasonable request.

	\bibliography{reference}

@article{annurev-conmatphys-031119-050605,
   author = "Kjaergaard, Morten and Schwartz, Mollie E. and Braumüller, Jochen and Krantz, Philip and Wang, Joel I.-J. and Gustavsson, Simon and Oliver, William D.",
   title = "Superconducting Qubits: Current State of Play", 
   journal= "Annual Review of Condensed Matter Physics",
   year = "2020",
   volume = "11",
   number = "Volume 11, 2020",
   pages = "369-395",
   doi = "https://doi.org/10.1146/annurev-conmatphys-031119-050605",
   url = "https://www.annualreviews.org/content/journals/10.1146/annurev-conmatphys-031119-050605",
   publisher = "Annual Reviews",
   issn = "1947-5462",
   type = "Journal Article",
  }

@article{PhysRevA.96.022330,
title = {Efficient $Z$ gates for quantum computing},
author = {McKay, David C. and Wood, Christopher J. and Sheldon, Sarah and Chow, Jerry M. and Gambetta, Jay M.},
journal = {Phys. Rev. A},
volume = {96},
issue = {2},
pages = {022330},
numpages = {8},
year = {2017},
month = {Aug},
publisher = {American Physical Society},
doi = {10.1103/PhysRevA.96.022330},
url = {https://link.aps.org/doi/10.1103/PhysRevA.96.022330}
}

@article{Kanaar2024,
doi = {10.1088/2058-9565/ad3d06},
url = {https://doi.org/10.1088/2058-9565/ad3d06},
year = {2024},
month = {apr},
publisher = {IOP Publishing},
volume = {9},
number = {3},
pages = {035011},
author = {Kanaar, David W and Kestner, J P},
title = {Neural-network-designed three-qubit gates robust against charge noise and crosstalk in silicon},
journal = {Quantum Sci. Technol.}
}

@article{PhysRevX.11.021058,
  title = {Realization of High-Fidelity CZ and $ZZ$-Free iSWAP Gates with a Tunable Coupler},
  author = {Sung, Youngkyu and Ding, Leon and Braum\"uller, Jochen and Veps\"al\"ainen, Antti and Kannan, Bharath and Kjaergaard, Morten and Greene, Ami and Samach, Gabriel O. and McNally, Chris and Kim, David and Melville, Alexander and Niedzielski, Bethany M. and Schwartz, Mollie E. and Yoder, Jonilyn L. and Orlando, Terry P. and Gustavsson, Simon and Oliver, William D.},
  journal = {Phys. Rev. X},
  volume = {11},
  issue = {2},
  pages = {021058},
  numpages = {32},
  year = {2021},
  month = {Jun},
  publisher = {American Physical Society},
  doi = {10.1103/PhysRevX.11.021058},
  url = {https://link.aps.org/doi/10.1103/PhysRevX.11.021058}
}

@article{10.1063/1.3691827,
    author = {Reich, Daniel M. and Ndong, Mamadou and Koch, Christiane P.},
    title = {Monotonically convergent optimization in quantum control using Krotov's method},
    journal = {The Journal of Chemical Physics},
    volume = {136},
    number = {10},
    pages = {104103},
    year = {2012},
    month = {03},
    issn = {0021-9606},
    doi = {10.1063/1.3691827},
    url = {https://doi.org/10.1063/1.3691827},
 
}

@Article{10.21468/SciPostPhys.7.6.080,
	title={{Krotov: A Python implementation of Krotov's method for quantum optimal control}},
	author={Michael H. Goerz and Daniel Basilewitsch and Fernando Gago-Encinas and Matthias G. Krauss and Karl P. Horn and Daniel M. Reich and Christiane P. Koch},
	journal={SciPost Phys.},
	volume={7},
	pages={080},
	year={2019},
	publisher={SciPost},
	doi={10.21468/SciPostPhys.7.6.080},
	url={https://scipost.org/10.21468/SciPostPhys.7.6.080},
}

@article{PhysRevA.68.062308,
  title = {Optimal control theory for unitary transformations},
  author = {Palao, Jos\'e P. and Kosloff, Ronnie},
  journal = {Phys. Rev. A},
  volume = {68},
  issue = {6},
  pages = {062308},
  numpages = {13},
  year = {2003},
  month = {Dec},
  publisher = {American Physical Society},
  doi = {10.1103/PhysRevA.68.062308},
  url = {https://link.aps.org/doi/10.1103/PhysRevA.68.062308}
}

@article{PhysRevApplied.18.064059,
  title = {Enhanced-Fidelity Ultrafast Geometric Quantum Computation Using Strong Classical Drives},
  author = {Chen, Ye-Hong and Miranowicz, Adam and Chen, Xi and Xia, Yan and Nori, Franco},
  journal = {Phys. Rev. Appl.},
  volume = {18},
  issue = {6},
  pages = {064059},
  numpages = {13},
  year = {2022},
  month = {Dec},
  publisher = {American Physical Society},
  doi = {10.1103/PhysRevApplied.18.064059},
  url = {https://link.aps.org/doi/10.1103/PhysRevApplied.18.064059}
}

@article{brif2010control,
  title = {Control of quantum phenomena: past,  present and future},
  volume = {12},
  ISSN = {1367-2630},
  url = {http://dx.doi.org/10.1088/1367-2630/12/7/075008},
  DOI = {10.1088/1367-2630/12/7/075008},
  number = {7},
  journal = {New. J. Phys.},
  publisher = {IOP Publishing},
  author = {Brif,  Constantin and Chakrabarti,  Raj and Rabitz,  Herschel},
  year = {2010},
  month = jul,
  pages = {075008}
}

@article{rabitz2000whither,
  title = {Whither the Future of Controlling Quantum Phenomena?},
  volume = {288},
  ISSN = {1095-9203},
  url = {http://dx.doi.org/10.1126/science.288.5467.824},
  DOI = {10.1126/science.288.5467.824},
  number = {5467},
  journal = {Science},
  publisher = {American Association for the Advancement of Science (AAAS)},
  author = {Rabitz,  Herschel and de Vivie-Riedle,  Regina and Motzkus,  Marcus and Kompa,  Karl},
  year = {2000},
  month = may,
  pages = {824–828}
}

@article{jirari2006quantum,
  title = {Quantum optimal control theory and dynamic coupling in the spin-boson model},
  author = {Jirari, H. and P\"otz, W.},
  journal = {Phys. Rev. A},
  volume = {74},
  issue = {2},
  pages = {022306},
  numpages = {18},
  year = {2006},
  month = {Aug},
  publisher = {American Physical Society},
  doi = {10.1103/PhysRevA.74.022306},
  url = {https://link.aps.org/doi/10.1103/PhysRevA.74.022306}
}

@article{werschnik2007quantum,
  title = {Quantum optimal control theory},
  volume = {40},
  ISSN = {1361-6455},
  url = {http://dx.doi.org/10.1088/0953-4075/40/18/R01},
  DOI = {10.1088/0953-4075/40/18/r01},
  number = {18},
  journal = {J. Phys. B},
  publisher = {IOP Publishing},
  author = {Werschnik,  J and Gross,  E K U},
  year = {2007},
  month = sep,
  pages = {R175–R211}
}

@article{chen2021shortcuts,
  title = {Shortcuts to Adiabaticity for the Quantum {Rabi} Model: Efficient Generation of Giant Entangled Cat States via Parametric Amplification},
  author = {Chen, Ye-Hong and Qin, Wei and Wang, Xin and Miranowicz, Adam and Nori, Franco},
  journal = {Phys. Rev. Lett.},
  volume = {126},
  issue = {2},
  pages = {023602},
  numpages = {8},
  year = {2021},
  month = {Jan},
  publisher = {American Physical Society},
  doi = {10.1103/PhysRevLett.126.023602},
  url = {https://link.aps.org/doi/10.1103/PhysRevLett.126.023602}
}

@article{chen2019fast,
  title = {Fast and high-fidelity generation of steady-state entanglement using pulse modulation and parametric amplification},
  author = {Chen, Ye-Hong and Qin, Wei and Nori, Franco},
  journal = {Phys. Rev. A},
  volume = {100},
  issue = {1},
  pages = {012339},
  numpages = {11},
  year = {2019},
  month = {Jul},
  publisher = {American Physical Society},
  doi = {10.1103/PhysRevA.100.012339},
  url = {https://link.aps.org/doi/10.1103/PhysRevA.100.012339}
}

@article{kang2018pulse,
  title = {Pulse design for multilevel systems by utilizing Lie transforms},
  author = {Kang, Yi-Hao and Chen, Ye-Hong and Shi, Zhi-Cheng and Huang, Bi-Hua and Song, Jie and Xia, Yan},
  journal = {Phys. Rev. A},
  volume = {97},
  issue = {3},
  pages = {033407},
  numpages = {13},
  year = {2018},
  month = {Mar},
  publisher = {American Physical Society},
  doi = {10.1103/PhysRevA.97.033407},
  url = {https://link.aps.org/doi/10.1103/PhysRevA.97.033407}
}

@article{shi2021two,
  title = {Two-level systems with periodic $N$-step driving fields: Exact dynamics and quantum state manipulations},
  author = {Shi, Zhi-Cheng and Chen, Ye-Hong and Qin, Wei and Xia, Yan and Yi, X. X. and Zheng, Shi-Biao and Nori, Franco},
  journal = {Phys. Rev. A},
  volume = {104},
  issue = {5},
  pages = {053101},
  numpages = {18},
  year = {2021},
  month = {Nov},
  publisher = {American Physical Society},
  doi = {10.1103/PhysRevA.104.053101},
  url = {https://link.aps.org/doi/10.1103/PhysRevA.104.053101}
}

@article{chen2016measuring,
  title = {Measuring and Suppressing Quantum State Leakage in a Superconducting Qubit},
  author = {Chen, Zijun and Kelly, Julian and Quintana, Chris and Barends,\textit{et al.}, R.},
  journal = {Phys. Rev. Lett.},
  volume = {116},
  issue = {2},
  pages = {020501},
  numpages = {5},
  year = {2016},
  month = {Jan},
  publisher = {American Physical Society},
  doi = {10.1103/PhysRevLett.116.020501},
  url = {https://link.aps.org/doi/10.1103/PhysRevLett.116.020501}
}

@article{PhysRevApplied.12.054023,
  title = {Suppression of Qubit Crosstalk in a Tunable Coupling Superconducting Circuit},
  author = {Mundada, Pranav and Zhang, Gengyan and Hazard, Thomas and Houck, Andrew},
  journal = {Phys. Rev. Appl.},
  volume = {12},
  issue = {5},
  pages = {054023},
  numpages = {10},
  year = {2019},
  month = {Nov},
  publisher = {American Physical Society},
  doi = {10.1103/PhysRevApplied.12.054023},
  url = {https://link.aps.org/doi/10.1103/PhysRevApplied.12.054023}
}

@article{malekakhlagh2020first,
  title = {First-principles analysis of cross-resonance gate operation},
  author = {Malekakhlagh, Moein and Magesan, Easwar and McKay, David C.},
  journal = {Phys. Rev. A},
  volume = {102},
  issue = {4},
  pages = {042605},
  numpages = {28},
  year = {2020},
  month = {Oct},
  publisher = {American Physical Society},
  doi = {10.1103/PhysRevA.102.042605},
  url = {https://link.aps.org/doi/10.1103/PhysRevA.102.042605}
}

@article{rol2019fast,
  title = {Fast, High-Fidelity Conditional-Phase Gate Exploiting Leakage Interference in Weakly Anharmonic Superconducting Qubits},
  author = {Rol, M. A. and Battistel, F. and Malinowski, F. K. and Bultink,\textit{et al.}, C. C.},
  journal = {Phys. Rev. Lett.},
  volume = {123},
  issue = {12},
  pages = {120502},
  numpages = {6},
  year = {2019},
  month = {Sep},
  publisher = {American Physical Society},
  doi = {10.1103/PhysRevLett.123.120502},
  url = {https://link.aps.org/doi/10.1103/PhysRevLett.123.120502}
}

@article{pioro2008electrically,
  title = {Electrically driven single-electron spin resonance in a slanting {Zeeman} field},
  volume = {4},
  ISSN = {1745-2481},
  url = {http://dx.doi.org/10.1038/nphys1053},
  DOI = {10.1038/nphys1053},
  number = {10},
  journal = {Nat. Phys.},
  publisher = {Springer Science and Business Media LLC},
  author = {Pioro-Ladrière,  M. and Obata,  T. and Tokura,  Y. and Shin,\textit{et al.},  Y.-S. },
  year = {2008},
  month = aug,
  pages = {776–779}
}

@article{10.1063/1.5089550,
    author = {Krantz, P. and Kjaergaard, M. and Yan, F. and Orlando, T. P. and Gustavsson, S. and Oliver, W. D.},
    title = {A quantum engineer's guide to superconducting qubits},
    journal = {Appl. Phys. Rev.},
    volume = {6},
    number = {2},
    pages = {021318},
    year = {2019},
    month = {06},
    issn = {1931-9401},
    doi = {10.1063/1.5089550},
    url = {https://doi.org/10.1063/1.5089550},
    
}

@article{annurev:/content/journals/10.1146/annurev-conmatphys-031119-050605,
   author = "Kjaergaard, Morten and Schwartz, Mollie E. and Braumüller, Jochen and Krantz, Philip and Wang, Joel I.-J. and Gustavsson, Simon and Oliver, William D.",
   title = "Superconducting Qubits: Current State of Play", 
   journal= "Annu. Rev. Condens. Matter Phys.",
   year = "2020",
   volume = "11",
   number = "Volume 11, 2020",
   pages = "369-395",
   doi = "https://doi.org/10.1146/annurev-conmatphys-031119-050605",
   url = "https://www.annualreviews.org/content/journals/10.1146/annurev-conmatphys-031119-050605",
   publisher = "Annual Reviews",
   issn = "1947-5462",
   type = "Journal Article",
  }

@article{PhysRevLett.133.033603,
  title = {Error-Tolerant Amplification and Simulation of the Ultrastrong-Coupling Quantum Rabi Model},
  author = {Chen, Ye-Hong and Shi, Zhi-Cheng and Nori, Franco and Xia, Yan},
  journal = {Phys. Rev. Lett.},
  volume = {133},
  issue = {3},
  pages = {033603},
  numpages = {7},
  year = {2024},
  month = {Jul},
  publisher = {American Physical Society},
  doi = {10.1103/PhysRevLett.133.033603},
  url = {https://link.aps.org/doi/10.1103/PhysRevLett.133.033603}
}

@article{khaneja2005optimal,
  title = {Optimal control of coupled spin dynamics: design of {NMR} pulse sequences by gradient ascent algorithms},
  volume = {172},
  ISSN = {1090-7807},
  url = {http://dx.doi.org/10.1016/j.jmr.2004.11.004},
  DOI = {10.1016/j.jmr.2004.11.004},
  number = {2},
  journal = {J. Magn. Reson.},
  publisher = {Elsevier BV},
  author = {Khaneja,  Navin and Reiss,  Timo and Kehlet,  Cindie and Schulte-Herbr\"{u}ggen,  Thomas and Glaser,  Steffen J.},
  year = {2005},
  month = feb,
  pages = {296–305}
}

@article{motzoi2009simple,
  title = {Simple Pulses for Elimination of Leakage in Weakly Nonlinear Qubits},
  author = {Motzoi, F. and Gambetta, J. M. and Rebentrost, P. and Wilhelm, F. K.},
  journal = {Phys. Rev. Lett.},
  volume = {103},
  issue = {11},
  pages = {110501},
  numpages = {4},
  year = {2009},
  month = {Sep},
  publisher = {American Physical Society},
  doi = {10.1103/PhysRevLett.103.110501},
  url = {https://link.aps.org/doi/10.1103/PhysRevLett.103.110501}
}

@article{carleo2019machine,
  title = {Machine learning and the physical sciences},
  author = {Carleo, Giuseppe and Cirac, Ignacio and Cranmer, Kyle and Daudet, Laurent and Schuld, Maria and Tishby, Naftali and Vogt-Maranto, Leslie and Zdeborov\'a, Lenka},
  journal = {Rev. Mod. Phys.},
  volume = {91},
  issue = {4},
  pages = {045002},
  numpages = {39},
  year = {2019},
  month = {Dec},
  publisher = {American Physical Society},
  doi = {10.1103/RevModPhys.91.045002},
  url = {https://link.aps.org/doi/10.1103/RevModPhys.91.045002}
}

@article{norambuena2024physics,
  title = {Physics-Informed Neural Networks for Quantum Control},
  author = {Norambuena, Ariel and Mattheakis, Marios and Gonz\'alez, Francisco J. and Coto, Ra\'ul},
  journal = {Phys. Rev. Lett.},
  volume = {132},
  issue = {1},
  pages = {010801},
  numpages = {7},
  year = {2024},
  month = {Jan},
  publisher = {American Physical Society},
  doi = {10.1103/PhysRevLett.132.010801},
  url = {https://link.aps.org/doi/10.1103/PhysRevLett.132.010801}
}

@article{Yang20241,
	title = {Mitigation of microwave crosstalk with parameterized single-qubit gate in superconducting quantum circuits},
	volume = {124},
	issn = {0003-6951},
	url = {https://doi.org/10.1063/5.0200014},
	doi = {10.1063/5.0200014},
	pages = {214001},
	number = {21},
	Journal = {Appl. Phys. Lett.},
	author = {Yang, Z. H. and Wang, Ruixia and Wang, Z. T. and Zhao, Peng and Huang, Kaixuan and Xu, Kai and Tian, Ye and Yu, H. F. and Zhao, S. P.},
	Year = {2024},
	
}

@article{Yang2024,
  Title = {Fast, universal scheme for calibrating microwave crosstalk in superconducting circuits},
  Volume = {125},
  ISSN = {1077-3118},
  url = {http://doi.org/10.1063/5.0211159},
  Doi = {10.1063/5.0211159},
  pages = {044001},
  Number = {4},
  Journal = {Appl. Phys. Lett.},
  Publisher = {AIP Publishing},
  Author = {Yang, Xiao-Yan and Zhang, Hai-Feng and Du, Lei and Tao,\textit{et al.}, Hao-Ran},
  Year = {2024},
  Month = jul 
}

@article{Wang2022,
  title = {Control and mitigation of microwave crosstalk effect with superconducting qubits},
  volume = {121},
  ISSN = {1077-3118},
  url = {http://dx.doi.org/10.1063/5.0115393},
  DOI = {10.1063/5.0115393},
  number = {15},
  pages = {152602},
  journal = {Appl. Phys. Lett.},
  publisher = {AIP Publishing},
  author = {Wang, Ruixia and Zhao, Peng and Jin, Yirong and Yu, Haifeng},
  year = {2022},
  month = oct 
}

@article{PhysRevLett.125.200504,
  title = {Suppression of Unwanted ${ZZ}$ Interactions in a Hybrid Two-Qubit System},
  author = {Ku, Jaseung and Xu, Xuexin and Brink, Markus and McKay, David C. and Hertzberg, Jared B. and Ansari, Mohammad H. and Plourde, B. L. T.},
  journal = {Phys. Rev. Lett.},
  volume = {125},
  issue = {20},
  pages = {200504},
  numpages = {6},
  year = {2020},
  month = {Nov},
  publisher = {American Physical Society},
  doi = {10.1103/PhysRevLett.125.200504},
  url = {https://link.aps.org/doi/10.1103/PhysRevLett.125.200504}
}

@article{PhysRevA.99.042327,
  title = {Learning robust and high-precision quantum controls},
  author = {Wu, Re-Bing and Ding, Haijin and Dong, Daoyi and Wang, Xiaoting},
  journal = {Phys. Rev. A},
  volume = {99},
  issue = {4},
  pages = {042327},
  numpages = {6},
  year = {2019},
  month = {Apr},
  publisher = {American Physical Society},
  doi = {10.1103/PhysRevA.99.042327},
  url = {https://link.aps.org/doi/10.1103/PhysRevA.99.042327}
}

@article{Dong2010,
  title = {Quantum control theory and applications: {A} survey},
  volume = {4},
  ISSN = {1751-8652},
  url = {http://dx.doi.org/10.1049/iet-cta.2009.0508},
  DOI = {10.1049/iet-cta.2009.0508},
  number = {12},
  journal = {IET Control Theory \& Applications},
  publisher = {Institution of Engineering and Technology (IET)},
  author = {Dong,  D. and Petersen,  I.R.},
  year = {2010},
  month = dec,
  pages = {2651–2671}
}

@article{Dong20160,
  title = {Learning robust pulses for generating universal quantum gates},
  volume = {6},
  pages = {2045-2322},
  url = {http://dx.doi.org/10.1038/srep36090},
  DOI = {10.1038/srep36090},
  number = {1},
  journal = {Sci. Rep.},
  publisher = {Springer Science and Business Media LLC},
  author = {Dong,  Daoyi and Wu,  Chengzhi and Chen,  Chunlin and Qi,  Bo and Petersen,  Ian R. and Nori,  Franco},
  year = {2016},
  month = oct 
}

@article{Dong20150,
  title = {Robust manipulation of superconducting qubits in the presence of fluctuations},
  volume = {5},
  Pages = {2045-2322},
  url = {http://dx.doi.org/10.1038/srep07873},
  DOI = {10.1038/srep07873},
  number = {1},
  journal = {Sci. Rep.},
  publisher = {Springer Science and Business Media LLC},
  author = {Dong,  Daoyi and Chen,  Chunlin and Qi,  Bo and Petersen,  Ian R. and Nori,  Franco},
  year = {2015},
  month = jan 
}

@article{PhysRevLett.130.043604,
  Title = {Quantum Coherent Control of a Single Molecular-Polariton Rotation},
  author = {Fan, Li-Bao and Shu, Chuan-Cun and Dong, Daoyi and He, Jun and Henriksen, Niels E. and Nori, Franco},
  Journal = {Phys. Rev. Lett.},
  Volume = {130},
  Issue = {4},
  Pages = {043604},
  Numpages = {8},
  Year = {2023},
  Month = {Jan},
  Publisher = {American Physical Society},
  Doi = {10.1103/PhysRevLett. 130.043604},
  url = {https://link.aps.org/doi/10.1103/PhysRevLett.130.043604}
}

@article{PhysRevResearch.7.L012049,
  Title = {Precise quantum control of molecular rotation toward a desired orientation},
  author = {Hong, Qian-Qian and Dong, Daoyi and Henriksen, Niels E. and Nori, Franco and He, Jun and Shu, Chuan-Cun},
  Journal = {Phys. Rev. Res.},
  Volume = {7},
  Issue = {1},
  Pages = {L 012049},
  Numpages = {10},
  Year = {2025},
  Month = {Feb},
  Publisher = {American Physical Society},
  Doi = {10.1103/PhysRevResearch. 7. L 012049},
  url = {https://link.aps.org/doi/10.1103/PhysRevResearch.7.L012049}
}

@article{PhysRevA.93.042307,
  title = {Crosstalk-insensitive method for simultaneously coupling multiple pairs of resonators},
  author = {Yang, Chui-Ping and Su, Qi-Ping and Zheng, Shi-Biao and Nori, Franco},
  journal = {Phys. Rev. A},
  volume = {93},
  issue = {4},
  pages = {042307},
  numpages = {9},
  year = {2016},
  month = {Apr},
  publisher = {American Physical Society},
  doi = {10.1103/PhysRevA.93.042307},
  url = {https://link.aps.org/doi/10.1103/PhysRevA.93.042307}
}

@ARTICLE{8759071,
  author={Dong, Daoyi and Xing, Xi and Ma, Hailan and Chen, Chunlin and Liu, Zhixin and Rabitz, Herschel},
  journal={IEEE Trans. Cybern.}, 
  title={Learning-Based Quantum Robust Control: Algorithm, Applications, and Experiments}, 
  year={2020},
  volume={50},
  number={8},
  pages={3581-3593},
  doi={10.1109/TCYB.2019.2921424}}

@article{PhysRevResearch.6.013142,
  title = {Mitigating crosstalk errors by randomized compiling: Simulation of the {BCS} model on a superconducting quantum computer},
  author = {Perrin, Hugo and Scoquart, Thibault and Shnirman, Alexander and Schmalian, J\"org and Snizhko, Kyrylo},
  journal = {Phys. Rev. Res.},
  volume = {6},
  issue = {1},
  pages = {013142},
  numpages = {24},
  year = {2024},
  month = {Feb},
  publisher = {American Physical Society},
  doi = {10.1103/PhysRevResearch.6.013142},
  url = {https://link.aps.org/doi/10.1103/PhysRevResearch.6.013142}
}

@article{PhysRevApplied.22.044072,
  title = {Crosstalk-robust quantum control in multimode bosonic systems},
  author = {You, Xinyuan and Lu, Yunwei and Kim, Taeyoon and K\"urk\ifmmode \mbox{\c{c}}\else \c{c}\fi{}\"uolu,\textit{et al.}, Doa Murat},
  journal = {Phys. Rev. Appl.},
  volume = {22},
  issue = {4},
  pages = {044072},
  numpages = {22},
  year = {2024},
  month = {Oct},
  publisher = {American Physical Society},
  doi = {10.1103/PhysRevApplied.22.044072},
  url = {https://link.aps.org/doi/10.1103/PhysRevApplied.22.044072}
}

@article{barends2014superconducting,
  title = {Superconducting quantum circuits at the surface code threshold for fault tolerance},
  volume = {508},
  ISSN = {1476-4687},
  url = {http://dx.doi.org/10.1038/nature13171},
  DOI = {10.1038/nature13171},
  number = {7497},
  journal = {Nature},
  publisher = {Springer Science and Business Media LLC},
  author = {Barends,  R. and Kelly,  J. and Megrant,  A. and Veitia,\textit{et al.},  A. },
  year = {2014},
  month = apr,
  pages = {500–503}
}

@article{bylander2011noise,
  title = {Noise spectroscopy through dynamical decoupling with a superconducting flux qubit},
  volume = {7},
  ISSN = {1745-2481},
  url = {http://dx.doi.org/10.1038/nphys1994},
  DOI = {10.1038/nphys1994},
  number = {7},
  journal = {Nat. Phys.},
  publisher = {Springer Science and Business Media LLC},
  author = {Bylander,  Jonas and Gustavsson,  Simon and Yan,  Fei and Yoshihara,\textit{et al.},  Fumiki},
  year = {2011},
  month = may,
  pages = {565–570}
}

@article{hocker2014characterization,
  title = {Characterization of control noise effects in optimal quantum unitary dynamics},
  author = {Hocker, David and Brif, Constantin and Grace, Matthew D. and Donovan, Ashley and Ho, Tak-San and Tibbetts, Katharine Moore and Wu, Rebing and Rabitz, Herschel},
  journal = {Phys. Rev. A},
  volume = {90},
  issue = {6},
  pages = {062309},
  numpages = {9},
  year = {2014},
  month = {Dec},
  publisher = {American Physical Society},
  doi = {10.1103/PhysRevA.90.062309},
  url = {https://link.aps.org/doi/10.1103/PhysRevA.90.062309}
}

@article{rudinger2021experimental,
  title = {Experimental Characterization of Crosstalk Errors with Simultaneous Gate Set Tomography},
  author = {Rudinger, Kenneth and Hogle, Craig W. and Naik, Ravi K. and Hashim,\textit{et al.}, Akel },
  journal = {PRX Quantum},
  volume = {2},
  issue = {4},
  pages = {040338},
  numpages = {21},
  year = {2021},
  month = {Nov},
  publisher = {American Physical Society},
  doi = {10.1103/PRXQuantum.2.040338},
  url = {https://link.aps.org/doi/10.1103/PRXQuantum.2.040338}
}

@article{yan2023calibration,
  title = {Calibration and cancellation of microwave crosstalk in superconducting circuits},
  volume = {32},
  ISSN = {1674-1056},
  url = {http://dx.doi.org/10.1088/1674-1056/acdc10},
  DOI = {10.1088/1674-1056/acdc10},
  number = {9},
  journal = {Chin. Phys. B},
  publisher = {IOP Publishing},
author = {Yan, Haisheng and Zhao, Shoukuan and Xiang, Zhongcheng and Wang, Ziting and Yang, Zhaohua and Xu, Kai and Tian, Ye and Yu, Haifeng and Zheng, Dongning and Fan, Heng and Zhao, Shiping},
  year = {2023},
  month = sep,
  pages = {094203}
}

@article{soare2014experimental,
  title = {Experimental noise filtering by quantum control},
  volume = {10},
  ISSN = {1745-2481},
  url = {http://dx.doi.org/10.1038/nphys3115},
  DOI = {10.1038/nphys3115},
  number = {11},
  journal = {Nat. Phys.},
  publisher = {Springer Science and Business Media LLC},
  author = {Soare,  A. and Ball,  H. and Hayes,  D. and Sastrawan,  J. and Jarratt,  M. C. and McLoughlin,  J. J. and Zhen,  X. and Green,  T. J. and Biercuk,  M. J.},
  year = {2014},
  month = oct,
  pages = {825–829}
}

@article{winick2021simulating,
  title = {Simulating and Mitigating Crosstalk},
  author = {Winick, Adam and Wallman, Joel J. and Emerson, Joseph},
  journal = {Phys. Rev. Lett.},
  volume = {126},
  issue = {23},
  pages = {230502},
  numpages = {6},
  year = {2021},
  month = {Jun},
  publisher = {American Physical Society},
  doi = {10.1103/PhysRevLett.126.230502},
  url = {https://link.aps.org/doi/10.1103/PhysRevLett.126.230502}
}

@article{tripathi2022suppression,
  title = {Suppression of Crosstalk in Superconducting Qubits Using Dynamical Decoupling},
  author = {Tripathi, Vinay and Chen, Huo and Khezri, Mostafa and Yip, Ka-Wa and Levenson-Falk, E.M. and Lidar, Daniel A.},
  journal = {Phys. Rev. Appl.},
  volume = {18},
  issue = {2},
  pages = {024068},
  numpages = {24},
  year = {2022},
  month = {Aug},
  publisher = {American Physical Society},
  doi = {10.1103/PhysRevApplied.18.024068},
  url = {https://link.aps.org/doi/10.1103/PhysRevApplied.18.024068}
}

@article{PhysRevA.84.022326,
  title = {Chopped random-basis quantum optimization},
  author = {Caneva, Tommaso and Calarco, Tommaso and Montangero, Simone},
  journal = {Phys. Rev. A},
  volume = {84},
  issue = {2},
  pages = {022326},
  numpages = {9},
  year = {2011},
  month = {Aug},
  publisher = {American Physical Society},
  doi = {10.1103/PhysRevA.84.022326},
  url = {https://link.aps.org/doi/10.1103/PhysRevA.84.022326}
}

@article{PhysRevLett.106.190501,
  title = {Optimal Control Technique for Many-Body Quantum Dynamics},
  author = {Doria, Patrick and Calarco, Tommaso and Montangero, Simone},
  journal = {Phys. Rev. Lett.},
  volume = {106},
  issue = {19},
  pages = {190501},
  numpages = {4},
  year = {2011},
  month = {May},
  publisher = {American Physical Society},
  doi = {10.1103/PhysRevLett.106.190501},
  url = {https://link.aps.org/doi/10.1103/PhysRevLett.106.190501}
}

@article{PhysRevLett.102.090401,
  title = {Optimal Control of a Qubit Coupled to a Non-Markovian Environment},
  author = {Rebentrost, P. and Serban, I. and Schulte-Herbr\"uggen, T. and Wilhelm, F. K.},
  journal = {Phys. Rev. Lett.},
  volume = {102},
  issue = {9},
  pages = {090401},
  numpages = {4},
  year = {2009},
  month = {Mar},
  publisher = {American Physical Society},
  doi = {10.1103/PhysRevLett.102.090401},
  url = {https://link.aps.org/doi/10.1103/PhysRevLett.102.090401}
}

@article{gungordu2022robust,
  title = {Robust quantum gates using smooth pulses and physics-informed neural networks},
  author = {G\"ung\"ord\"u, Utkan and Kestner, J. P.},
  journal = {Phys. Rev. Res.},
  volume = {4},
  issue = {2},
  pages = {023155},
  numpages = {9},
  year = {2022},
  month = {May},
  publisher = {American Physical Society},
  doi = {10.1103/PhysRevResearch.4.023155},
  url = {https://link.aps.org/doi/10.1103/PhysRevResearch.4.023155}
}

@article{PhysRevApplied.12.064022,
  title = {Methods for Measuring Magnetic Flux Crosstalk between Tunable Transmons},
  author = {Abrams, Deanna M. and Didier, Nicolas and Caldwell, Shane A. and Johnson, Blake R. and Ryan, Colm A.},
  journal = {Phys. Rev. Appl.},
  volume = {12},
  issue = {6},
  pages = {064022},
  numpages = {11},
  year = {2019},
  month = {Dec},
  publisher = {American Physical Society},
  doi = {10.1103/PhysRevApplied.12.064022},
  url = {https://link.aps.org/doi/10.1103/PhysRevApplied.12.064022}
}

@article{Huang2020,
  title = {Superconducting quantum computing: a review},
  volume = {63},
  ISSN = {1869-1919},
  url = {http://dx.doi.org/10.1007/s11432-020-2881-9},
  DOI = {10.1007/s11432-020-2881-9},
  number = {8}, 
  journal = {Sci. China Inf. Sci.}, 
  publisher = {Springer Science and Business Media LLC},
  author = {Huang, H.-L. and Wu, D. and Fan, D. and Zhu, X.}, 
  year = {2020},
  month = jul,
  pages = {180501}
}

@article{Sarovar2020,
  title = {Detecting crosstalk errors in quantum information processors},
  volume = {4},
  ISSN = {2521-327X},
  url = {http://dx.doi.org/10.22331/q-2020-09-11-321},
  DOI = {10.22331/q-2020-09-11-321},
  journal = {Quantum},
  publisher = {Verein zur Forderung des Open Access Publizierens in den Quantenwissenschaften},
  author = {Sarovar,  Mohan and Proctor,  Timothy and Rudinger,  Kenneth and Young,  Kevin and Nielsen,  Erik and Blume-Kohout,  Robin},
  year = {2020},
  month = sep,
  pages = {321}
}

@article{PhysRevA.83.012308,
  title = {Analytic control methods for high-fidelity unitary operations in a weakly nonlinear oscillator},
  author = {Gambetta, J. M. and Motzoi, F. and Merkel, S. T. and Wilhelm, F. K.},
  journal = {Phys. Rev. A},
  volume = {83},
  issue = {1},
  pages = {012308},
  numpages = {13},
  year = {2011},
  month = {Jan},
  publisher = {American Physical Society},
  doi = {10.1103/PhysRevA.83.012308},
  url = {https://link.aps.org/doi/10.1103/PhysRevA.83.012308}
}

@article{Tanttu2024,
  title = {Assessment of the errors of high-fidelity two-qubit gates in silicon quantum dots},
  volume = {20},
  ISSN = {1745-2481},
  url = {http://dx.doi.org/10.1038/s41567-024-02614-w},
  DOI = {10.1038/s41567-024-02614-w},
  number = {11},
  journal = {Nat. Phys.},
  publisher = {Springer Science and Business Media LLC},
  author = {Tanttu,  Tuomo and Lim,  Wee Han and Huang,  Jonathan Y. and Dumoulin Stuyck,  Nard and Gilbert,  Will and Su,  Rocky Y. and Feng,  MengKe and others},
  year = {2024},
  month = aug,
  pages = {1804–1809}
}

@article{PhysRevX.3.041013,
  Title = {Error Suppression and Error Correction in Adiabatic Quantum Computation: Techniques and Challenges},
  author = {Young, Kevin C. and Sarovar, Mohan and Blume-Kohout, Robin},
  Journal = {Phys. Rev. X},
  Volume = {3},
  Issue = {4},
  Pages = {041013},
  Numpages = {13},
  Year = {2013},
  Month = {Nov},
  Publisher = {American Physical Society},
  Doi = {10.1103/PhysRevX.3.041013},
  url = {https://link.aps.org/doi/10.1103/PhysRevX.3.041013}
}

@article{PhysRevLett.111.030405,
  title = {Optimal Coherent Control to Counteract Dissipation},
  author = {Sauer, Simeon and Gneiting, Clemens and Buchleitner, Andreas},
  journal = {Phys. Rev. Lett.},
  volume = {111},
  issue = {3},
  pages = {030405},
  numpages = {5},
  year = {2013},
  month = {Jul},
  publisher = {American Physical Society},
  doi = {10.1103/PhysRevLett.111.030405},
  url = {https://link.aps.org/doi/10.1103/PhysRevLett.111.030405}
}

@article{Cheng2023,
  title = {Noisy intermediate-scale quantum computers},
  volume = {18},
  ISSN = {2095-0470},
  url = {http://dx.doi.org/10.1007/s11467-022-1249-z},
  DOI = {10.1007/s11467-022-1249-z},
  number = {2},
  journal = {Front. Phys.},
  publisher = {China Engineering Science Press Co. Ltd.},
  author = {Cheng,  Bin and Deng,  Xiu-Hao and Gu,  Xiu and He,  Yu and others},
  year = {2023},
  month = mar,
  pages = {21308}
}

@article{PhysRevB.74.172505,
  title = {Switchable resonant coupling of flux qubits},
  author = {Grajcar, M. and Liu, Yu-xi and Nori, Franco and Zagoskin, A. M.},
  journal = {Phys. Rev. B},
  volume = {74},
  issue = {17},
  pages = {172505},
  numpages = {4},
  year = {2006},
  month = {Nov},
  publisher = {American Physical Society},
  doi = {10.1103/PhysRevB.74.172505},
  url = {https://link.aps.org/doi/10.1103/PhysRevB.74.172505}
}

@article{deGroot2012,
  title = {Selective darkening of degenerate transitions for implementing quantum controlled-{NOT} gates},
  volume = {14},
  ISSN = {1367-2630},
  url = {http://dx.doi.org/10.1088/1367-2630/14/7/073038},
  DOI = {10.1088/1367-2630/14/7/073038},
  number = {7},
  journal = {New J. Phys.},
  publisher = {IOP Publishing},
  author = {de Groot,  P C and Ashhab,  S and Lupaşcu,  A and DiCarlo,  L and Nori,  Franco and Harmans,  C J P M and Mooij,  J E},
  year = {2012},
  month = jul,
  pages = {073038}
}

@article{PhysRevB.77.014510,
  title = {Interqubit coupling mediated by a high-excitation-energy quantum object},
  author = {Ashhab, S. and Niskanen, A. O. and Harrabi, K. and Nakamura, Y. and Picot, T. and de Groot, P. C. and Harmans, C. J. P. M. and Mooij, J. E. and Nori, Franco},
  journal = {Phys. Rev. B},
  volume = {77},
  issue = {1},
  pages = {014510},
  numpages = {13},
  year = {2008},
  month = {Jan},
  publisher = {American Physical Society},
  doi = {10.1103/PhysRevB.77.014510},
  url = {https://link.aps.org/doi/10.1103/PhysRevB.77.014510}
}

@article{PhysRevLett.96.067003,
  title = {Controllable Coupling between Flux Qubits},
  author = {Liu, Yu-xi and Wei, L. F. and Tsai, J. S. and Nori, Franco},
  journal = {Phys. Rev. Lett.},
  volume = {96},
  issue = {6},
  pages = {067003},
  numpages = {4},
  year = {2006},
  month = {Feb},
  publisher = {American Physical Society},
  doi = {10.1103/PhysRevLett.96.067003},
  url = {https://link.aps.org/doi/10.1103/PhysRevLett.96.067003}
}

@article{PhysRevB.76.132513,
  title = {Switchable coupling for superconducting qubits using double resonance in the presence of crosstalk},
  author = {Ashhab, S. and Nori, Franco},
  journal = {Phys. Rev. B},
  volume = {76},
  issue = {13},
  pages = {132513},
  numpages = {4},
  year = {2007},
  month = {Oct},
  publisher = {American Physical Society},
  doi = {10.1103/PhysRevB.76.132513},
  url = {https://link.aps.org/doi/10.1103/PhysRevB.76.132513}
}

@article{refId0,
	author = {Koch, Christiane P and Boscain, Ugo and Calarco, Tommaso and Dirr, Gunther and Filipp, Stefan and Glaser, Steffen J and Kosloff, Ronnie and Montangero, Simone and Schulte-Herbr{\"u}ggen, Thomas and Sugny, Dominique and others},
	title = {Quantum optimal control in quantum technologies. Strategic report on current status, visions and goals for research in Europe},
	DOI= "10.1140/epjqt/s40507-022-00138-x",
	url= "https://doi.org/10.1140/epjqt/s40507-022-00138-x",
	journal = {EPJ Quantum Technol.},
	year = 2022,
	volume = 9,
	number = 1,
	pages = "19",
}

@article{PhysRevB.74.184504,
  title = {Generalized switchable coupling for superconducting qubits using double resonance},
  author = {Ashhab, S. and Matsuo, Shigemasa and Hatakenaka, Noriyuki and Nori, Franco},
  journal = {Phys. Rev. B},
  volume = {74},
  issue = {18},
  pages = {184504},
  numpages = {6},
  year = {2006},
  month = {Nov},
  publisher = {American Physical Society},
  doi = {10.1103/PhysRevB.74.184504},
  url = {https://link.aps.org/doi/10.1103/PhysRevB.74.184504}
}

@article{wendin2017quantum,
  title = {Quantum information processing with superconducting circuits: a review},
  volume = {80},
  ISSN = {1361-6633},
  url = {http://dx.doi.org/10.1088/1361-6633/aa7e1a},
  DOI = {10.1088/1361-6633/aa7e1a},
  number = {10},
  journal = {Rep. Prog. Phys.},
  publisher = {IOP Publishing},
  author = {Wendin,  G},
  year = {2017},
  month = sep,
  pages = {106001}
}

@article{Li2022,
  title = {Pulse-level noisy quantum circuits with QuTiP},
  volume = {6},
  ISSN = {2521-327X},
  url = {http://dx.doi.org/10.22331/q-2022-01-24-630},
  DOI = {10.22331/q-2022-01-24-630},
  journal = {Quantum},
  publisher = {Verein zur Forderung des Open Access Publizierens in den Quantenwissenschaften},
  author = {Li,  Boxi and Ahmed,  Shahnawaz and Saraogi,  Sidhant and Lambert,  Neill and Nori,  Franco and Pitchford,  Alexander and Shammah,  Nathan},
  year = {2022},
  month = jan,
  pages = {630}
}

@article{zhao2022quantum,
  title = {Quantum Crosstalk Analysis for Simultaneous Gate Operations on Superconducting Qubits},
  author = {Zhao, Peng and Linghu, Kehuan and Li, Zhiyuan and Xu, Peng and Wang, Ruixia and Xue, Guangming and Jin, Yirong and Yu, Haifeng},
  journal = {PRX Quantum},
  volume = {3},
  issue = {2},
  pages = {020301},
  numpages = {21},
  year = {2022},
  month = {Apr},
  publisher = {American Physical Society},
  doi = {10.1103/PRXQuantum.3.020301},
  url = {https://link.aps.org/doi/10.1103/PRXQuantum.3.020301}
}

@article{zhou2023quantum,
  title = {Quantum Crosstalk Robust Quantum Control},
  author = {Zhou, Zeyuan and Sitler, Ryan and Oda, Yasuo and Schultz, Kevin and Quiroz, Gregory},
  journal = {Phys. Rev. Lett.},
  volume = {131},
  issue = {21},
  pages = {210802},
  numpages = {7},
  year = {2023},
  month = {Nov},
  publisher = {American Physical Society},
  doi = {10.1103/PhysRevLett.131.210802},
  url = {https://link.aps.org/doi/10.1103/PhysRevLett.131.210802}
}
\end{document}